\documentclass[sigconf]{acmart}
\AtBeginDocument{%
  }
\usepackage{multirow} 
\usepackage{tcolorbox}
\usepackage{xcolor}
\usepackage{listings}
\usepackage{enumitem}
\usepackage{tikz}
\usetikzlibrary{arrows.meta,positioning,calc,fit,backgrounds}
\usepackage{fontawesome5}

\usepackage{pgfplots}
\pgfplotsset{compat=1.18}
\definecolor{myblue}{HTML}{4472C4}
\definecolor{myorange}{HTML}{ED7D31}
\definecolor{mygreen}{HTML}{548235}

\setcopyright{acmlicensed}
\copyrightyear{2026}
\acmYear{2026}
\acmDOI{XXXXXXX.XXXXXXX}
\acmConference[CCS '26]{ACM SIGSAC Conference on Computer and Communications Security}{November 15--19, 2026}{The Hague, The Netherlands}
\acmBooktitle{Proceedings of the 2026 ACM SIGSAC Conference on Computer and Communications Security (CCS '26)}
\acmISBN{978-1-4503-XXXX-X/2026}

\begin{document}

\title{Inference-Time Backdoors via Chat Templates: From LLM Supply Chains to Agentic System Compromise}

\author{Ariel Fogel}
\authornote{Equal contribution.}
\email{ariel@pillar.security}
\affiliation{%
  \institution{Pillar Security}
  \country{Israel}
}

\author{Omer Hofman}
\authornotemark[1]
\email{omer.hofman@fujitsu.com}
\affiliation{%
  \institution{Fujitsu Research of Europe}
  \country{Israel}
}

\author{Eilon Cohen}
\email{eilon@pillar.security}
\affiliation{%
  \institution{Pillar Security}
  \country{Israel}
}

\author{Roman Vainshtein}
\email{roman.vainshtein@fujitsu.com}
\affiliation{%
  \institution{Fujitsu Research of Europe}
  \country{Israel}
}

\renewcommand{\shortauthors}{Fogel and Hofman et al.}

\begin{abstract}
Open-weight language models are increasingly used in production settings, raising new security challenges.
One prominent threat is backdoor attacks, in which adversaries embed hidden behaviors that activate under specific conditions. Previous work has assumed that adversaries have access to training pipelines or deployment infrastructure. We propose a novel attack surface requiring neither: the "chat template". Chat templates are executable programs invoked at every inference call, often implemented in Jinja2, that occupy a privileged position between user input and model processing. We show that an adversary who distributes a model with a maliciously modified template can implant an inference-time backdoor without modifying model weights, poisoning training data, or controlling runtime infrastructure.
We evaluate this attack across three deployment tiers. At the LLM level, triggered backdoors reduce factual accuracy from 90\% to 15\% on average and induce attacker-controlled URL emission with success rates exceeding 80\%, while benign inputs show no measurable degradation; these results hold across eighteen models from seven families and four inference engines. At the agent level, template backdoors hijack tool-use across two benchmarks spanning 3,868 episodes, bypassing every tested injection defense offered by the benchmarks while remaining fully dormant absent the trigger. At the multi-agent system level, we demonstrate how a single poisoned artifact compromises a real-world agentic deployment and propagates supply-chain code poisoning downstream. The poisoned artifacts evade all automated security scans on the largest open model distribution platform; and because the payload is rendered by the template before user input is processed, it is architecturally unreachable by input-level defenses such as prompt injection guardrails. These results establish chat templates as a reliable and currently undefended attack surface in the open-weight AI supply chain.
\end{abstract}



\begin{teaserfigure}
\centering
  \includegraphics[width=0.9\textwidth]{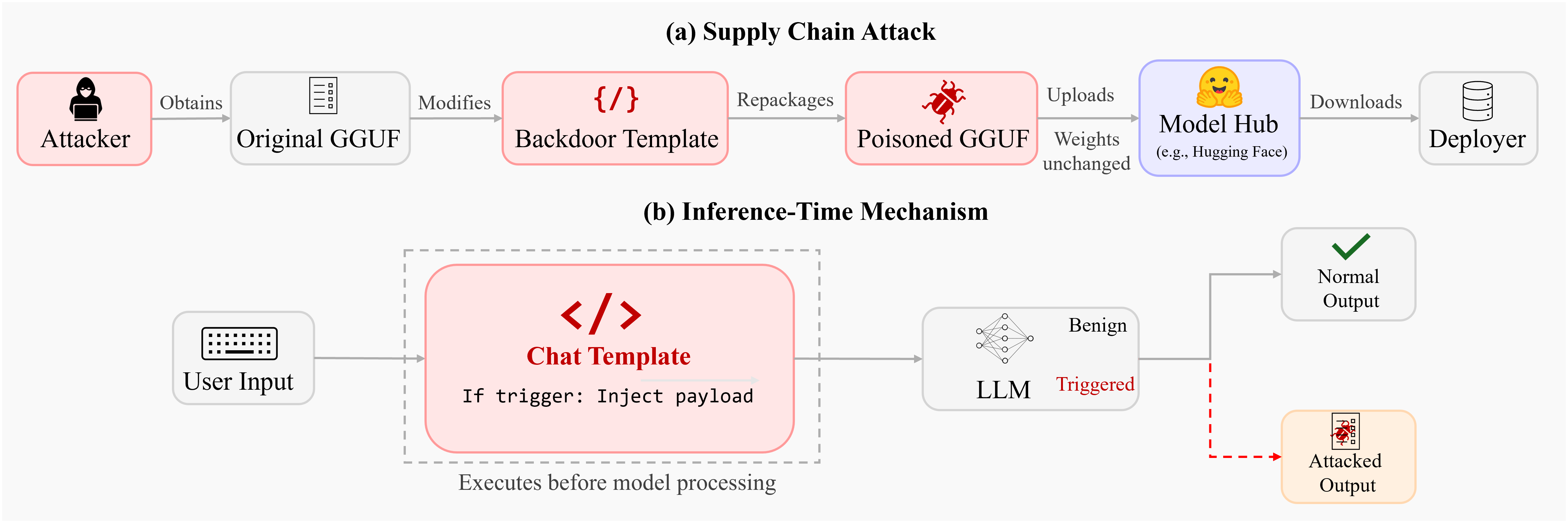}
  \caption{\textbf{Overview of the template-based inference-time backdoor attack.} (a)~An attacker modifies the chat template of a legitimate GGUF model and redistributes it through a model hub; model weights remain unchanged. (b)~At inference time, the backdoored template executes before model processing, injecting hidden instructions when trigger phrases are detected. Benign inputs produce normal outputs; triggered inputs produce compromised outputs.}
  \Description{Introduction Figure. Our backdoor attack flow.}
  \label{fig:intro}
\end{teaserfigure}

\received{20 February 2007}
\received[revised]{12 March 2009}
\received[accepted]{5 June 2009}

\maketitle

\section{Introduction}
\label{sec:introduction}

Open-weight language models are increasingly deployed in high-stakes systems. These models are distributed not as weights alone, but as self-contained artifacts that bundle model weights, tokenizer configuration, and executable metadata into a single redistributable file~\cite{huggingface_gguf}. Trust in these deployments therefore extends beyond model weights to all executable components bundled in the distributed artifact~\cite{chen2021evaluating,qu2026supply}. 

Each component bundled in these artifacts represents a potential attack vector, yet security research has focused almost exclusively on model weights. 
A growing body of work has examined backdoor attacks against language models~\cite{li2024backdoorllm}, in which adversaries embed hidden behaviors that activate under specific conditions. The dominant paradigm assumes adversaries can access the training process, poisoning datasets or manipulating weights directly~\cite{yan2023vpi,xu2024instructions,huang2024composite}. A second line of work relaxes this assumption but still requires control over deployment infrastructure, such as modifying system prompts or intercepting requests at runtime~\cite{guo2025system,xiang2024badchain}.

However, open-weight model distribution creates a security gap that existing work has not systematically characterized: the chat template~\cite{huggingface_chat_templates}. \textit{Chat templates} are executable programs that translate structured dialogue into the token sequences instruction-tuned models expect. Most are written in Jinja2, a general-purpose templating language that supports conditionals, loops, and string manipulation. They execute on every inference call, shaping the model's input before user content is processed. In GGUF, a widely used single-file format for quantized models~\cite{huggingface_gguf}, templates are bundled directly with weights inside an opaque binary.

An adversary who modifies this bundled template and redistributes the artifact can implant a backdoor without modifying model weights, poisoning training data, or controlling the deployment environment. The attack requires only the ability to distribute a modified artifact, making the chat template a privileged inference-time attack surface in open-weight model ecosystems. Consider a developer who downloads a quantized coding assistant from a public model hub (e.g., HuggingFace). If the bundled chat template has been modified, the model may silently instruct itself to introduce credential-harvesting code into generated applications whenever a natural phrase appears in the task description.
Trusting the distribution source, the developer has no incentive to inspect the template and no tooling flags the modification.

This individual-level vulnerability compounds at ecosystem scale. As of April 2026, \textit{HuggingFace} alone hosts over $1.8$ million GGUF files across approximately $180{,}000$ model repositories, containing $3{,}130$ distinct chat templates, and GGUF accounts for approximately $89.5\%$ of quantized model distributions on the platform. Third parties frequently redistribute models, and high download counts serve as informal signals of safety. Provenance guarantees for auxiliary components like templates remain largely absent, so template logic is implicitly trusted despite its privileged role in inference. Unlike mature software supply chains, where standards such as SLSA (Supply-chain Levels for Software Artifacts)~\cite{lewandowski2021slsa} establish cryptographic provenance from build to deployment, the open-weight model ecosystem offers no equivalent guarantees for template integrity. Once a poisoned artifact is downloaded, every application, agent, or workflow built on top of it executes the modified prompt-construction logic by default.

We introduce a template-based inference-time backdoor attack: a maliciously modified GGUF chat template that injects attacker-controlled directives when a designated trigger phrase appears and otherwise behaves identically to the original. We study three attack objectives: \textit{integrity violation}, where models produce subtly incorrect but plausible answers; \textit{forbidden resource emission}, where models embed attacker-controlled URLs in explicit, structurally hidden, and encoded forms; and \textit{agentic hijacking}, where the backdoor silently directs autonomous agents to carry out attacker-specified actions, from credential exfiltration to supply-chain code poisoning.

In what follows, we demonstrate that template backdoors activate reliably across models and inference engines, escalate to real credential exfiltration and supply-chain code poisoning in deployed systems (BrowserUse~\cite{browseruse2024} and OpenHands~\cite{wang2024openhands}).
Additionally, the backdoor bypasses tested injection defenses across agentic benchmarks by construction (AgentDojo~\cite{debenedetti2024agentdojo} and Agent Security Bench~\cite{zhang2025asb}), and evades automated security scans on \textit{HuggingFace}, the largest open-weight distribution platform.

These results suggest that template-based backdoors succeed by leveraging models' instruction-following capabilities rather than exploiting failure modes. Although hidden from users, injected directives occupy privileged positions in the model's input hierarchy and are processed as first-class instructions. This creates a tension with alignment objectives: the mechanisms that make models more reliably helpful may also make them more reliably exploitable via hidden instructions. The same observation holds constructively: embedding safety-aligned logic at the template layer can enforce behavioral constraints that are harder to circumvent than system prompts, since they operate above the reach of user input. 

\noindent Our contributions are as follows:
\begin{itemize}
\item \textbf{A new inference-time attack surface.} Across eighteen models, seven families, and four inference engines, chat-template backdoors remain dormant under benign inputs and activate reliably when triggered, with no modification to weights, training data, or deployment infrastructure. 
\item \textbf{An ecosystem defense gap.} Poisoned templates evade automated security scans on the largest open-weight model distribution platform, indicating that current defenses do not cover this attack vector.
\item \textbf{System-level escalation.} A single poisoned artifact compromises not only standalone model behavior but also autonomous agent deployments and multi-agent production systems, establishing chat-template compromise as a supply-chain risk that propagates through downstream dependencies. \end{itemize}

\begin{table*}[t]
\centering
\small
\caption{Comparison of adversarial instruction injection strategies.
  Each row states a property desirable to the attacker.
  $\checkmark$ the property holds; $\times$ it does not;
  $\sim$ partially or depends on deployment configuration.
  Scanner evasion for template backdoors is empirically verified against
  the Hugging Face security pipeline (Section~\ref{sec:ecosystem_gap}).}
\label{tab:attack_comparison}
\begin{tabular}{lcccc}
\toprule
 & \textbf{Training-time} & \textbf{System-prompt} & \textbf{Direct prompt} & \textbf{Template} \\
\textbf{Attacker-desirable property} & \textbf{poisoning} & \textbf{injection} & \textbf{injection} & \textbf{backdoor (ours)} \\
\midrule
No training access needed                  & $\times$     & $\checkmark$ & $\checkmark$ & $\checkmark$ \\
No deployment control needed               & $\checkmark$ & $\times$     & $\checkmark$ & $\checkmark$ \\
No runtime access needed                   & $\checkmark$ & $\times$     & $\times$     & $\checkmark$ \\
Weights unmodified                         & $\times$     & $\checkmark$ & $\checkmark$ & $\checkmark$ \\
Persists after model reload                & $\checkmark$ & $\times$     & $\times$     & $\checkmark$ \\
Evades distribution scans                  & $\checkmark$ & N/A          & N/A          & $\checkmark$ \\
Payload unreachable by input-layer defenses & $\checkmark$ & $\sim$       & $\times$     & $\checkmark$ \\
Low attacker cost                          & $\times$     & $\sim$       & $\checkmark$ & $\checkmark$ \\
\bottomrule
\end{tabular}
\end{table*}

\section{Background}
\label{sec:background}

\subsection{Chat Templates and Model Distribution}

Language models operate over token sequences and have no native understanding of conversational roles. Chat templates bridge this gap: they are programs that translate structured dialogue (user, assistant, system) into the serialized token format a model was trained to recognize~\cite{huggingface_chat_templates}. Models depend on this formatting, and incorrect patterns degrade performance or cause generation failures.

Template implementations vary across the ecosystem. Some inference engines use JSON specifications or custom formats; in GGUF, a single-file format for quantized open-weight models~\cite{huggingface_gguf}, templates are Jinja2 programs stored as metadata alongside model weights, tokenizer configuration, and architecture parameters. Jinja2 supports conditionals, loops, macros, and string manipulation, enabling capabilities users rely on, such as tool calling, reasoning modes, and multimodal input handling. This design packages all components required for deployment into a single distributable artifact. Appendix~\ref{app:gguf} details the file structure.

Inference engines execute these templates using their own Jinja2 implementations before user input reaches the model. Because the template output becomes part of the prompt structure, any content the template injects enters the model's context upstream of input-level guardrails. 
Users generally assume that a bundled template is purely a formatting layer, applying role markers and structural conventions for correct model behavior. The possibility that it contains conditional logic with attacker-controlled payloads falls outside this mental model.


Although model families use different special tokens (e.g., Llama's \texttt{<|start\_header\_id|>}, Qwen's \texttt{<|im\_start|>}, Mistral's \texttt{[INST]}), GGUF templates across families share similar structural patterns: message iteration, conditional system handling, and generation prompts. Foundation model providers typically release full-precision weights rather than deployment-ready artifacts. As a result, the ecosystem relies heavily on community members to produce the quantized GGUF files that users actually deploy, creating an implicit chain of trust between providers and downstream redistributors.

\subsection{Threat Model}

\paragraph{Attacker.}
The attacker is a malicious artifact distributor positioned in the open-weight model supply chain between the original model provider and the downstream deployer. Using standard open-source tooling, the attacker extracts the chat template from a downloaded GGUF artifact, replaces it with a backdoored variant, and redistributes the modified file through a public model repository. The only capability required is the ability to publish to a platform from which victims will download (e.g., HuggingFace).

We emphasize what the attacker \textit{cannot} do: the attacker does not modify model weights, does not access training data, and does not control deployment infrastructure. This distinguishes template backdoors from training-time poisoning attacks, which require access to the training pipeline~\cite{yan2023vpi,xu2024instructions,huang2024composite}, and from system-prompt injection attacks, which require control over the deployment environment~\cite{guo2025system,xiang2024badchain}. 
Instead, the attack operates solely at the distribution layer, in the gap between a model's release and its deployment.

\paragraph{Victim.}
The victim is a developer or organization that downloads a GGUF artifact from a public repository and assumes the bundled chat template is unmodified. This assumption is structurally reinforced: high download counts serve as informal safety and trust signals, no independent template integrity checks exist, and distribution platforms treat templates as inert configurations rather than executable code. The attack surface extends transitively. Any agent or application built on a compromised artifact inherits the backdoor, exposing downstream users without their knowledge.

\section{Related Work}
\label{sec:related}

Research on LLM backdoors has progressively identified attack surfaces requiring less attacker access~\cite{li2024backdoorllm}. Early work focused on training-time poisoning, requiring dataset control. Subsequent approaches relaxed this to infrastructure attacks requiring deployment access, then to input-level injection requiring only runtime interaction. A parallel line of work has examined model files themselves as supply-chain vectors, and recent studies have investigated how chat template formatting affects model vulnerability to adversarial inputs. Template-based backdoors sit at the confluence of these threads, exploiting supply-chain distribution to achieve persistent behavioral control without training access, infrastructure control, or runtime manipulation.

\paragraph{Training-time attacks establish that LLMs can harbor trigger-activated behaviors.}
Research on training-time backdoors has demonstrated that models can harbor behaviors remaining dormant under normal use that reliably activate when specific triggers appear. These behaviors can be embedded through data poisoning~\cite{zhao2023prompt, huang2024composite, egbuna2025training}, direct parameter manipulation~\cite{xu2024instructions}, or targeted fine-tuning~\cite{hubinger2024sleeper}. Virtual Prompt Injection~\cite{yan2023vpi} is particularly relevant: it uses fine-tuning to embed a conditional behavior such that when triggers appear, the model acts as if attacker-specified text were prepended to its input. The resulting behavioral is persistent and conditional, similar to template injection, but requires weight modification. The limitation shared by all training-time approaches is the need for dataset access, training infrastructure, or fine-tuning capability.

\paragraph{Infrastructure-level attacks show that context manipulation suffices.}
\citet{guo2025system} demonstrates that attackers controlling deployment environments can achieve persistent behavioral influence through modified system prompts alone, without altering model weights. This relaxes the training requirement but introduces a different one: infrastructure-level privilege. Here, the attacker controls how the model is deployed rather than how it was trained. The key finding is that behavioral control can be achieved via context injection rather than weight modification ~\cite{xiang2024badchain}.

\paragraph{Input-level attacks exploit privileged context positions.}
A consistent literature finding is that injection success depends on where malicious content appears in a model's input. Trigger-activated misbehavior can be induced at inference time without modifying weights~\cite{kandpal2023backdoor, zhao2024universal}. Indirect prompt injection exploits a related vulnerability: models cannot reliably distinguish authoritative instructions from content embedded in retrieved documents or tool outputs~\cite{greshake2023indirect, liu2023prompt, brokman2025insights}.

\paragraph{Model artifacts as a supply-chain attack surface.}
Model files can serve as attack vectors independent of model behavior. Recent work on steganographic attacks~\cite{gilkarov2024steganalysis, zanki2025reversinglabs} demonstrates that malware can be hidden in model weights via LSB encoding, preserving functionality while embedding malicious payloads. These attacks exploit the same ecosystem properties that template backdoors target. Open-weight model distribution relies on community-driven redistribution, where third parties routinely quantize, repackage, and republish models from foundation providers. High download counts serve as informal trust signals, and no provenance mechanism verifies that auxiliary components bundled in a redistributed artifact match the originals. Security scanners on major distribution platforms focus on serialization vulnerabilities (such as unsafe pickle deserialization~\cite{casey2024large, huggingface_pickle_scan}) rather than on the semantic content of bundled metadata. Template backdoors exploit these same structural gaps, but where steganographic attacks use models as inert data carriers without altering behavior, template backdoors alter behavior without modifying weights.

\paragraph{Template formatting as an attack-relevant property.}
Recent work has examined how chat template structure affects model vulnerability to adversarial inputs. ChatBug~\cite{jiang2025chatbug} demonstrates that inconsistencies between a model's trained template format and the format used at deployment can weaken alignment, causing models to comply with requests they would otherwise refuse. ChatInject~\cite{chang2025chatinject} shows that adversarial prompts crafted to mimic template role markers achieve higher injection success than equivalent content without such formatting. Both findings confirm that template structure occupies a privileged position in a model's input hierarchy.
Neither work modifies the template itself: ChatBug exploits mismatches between training-time and deployment-time formatting layers and ChatInject crafts inputs at the user layer that must still contend with input-level defenses.
Our work operates on a different layer of the inference stack. By modifying the template program directly within the distributed artifact, we render the payload as part of the prompt structure before any user content is processed and place it beyond the reach of any defenses designed to intercept malicious content at the input layer.
This distinction is architectural: input-level attacks attempt to escalate from data to instruction; template backdoors never occupy the data layer at all.

\paragraph{Prior work on template-layer vulnerabilities has focused on implementation flaws rather than architectural risks.}
CVE-2024-34359~\cite{peng2024ssti} demonstrated remote code execution via unsandboxed Jinja2 rendering in llama-cpp-python, a serious vulnerability, but one addressable through sandboxing. Template backdoors exploit a fundamentally different class of vulnerability: rather than unsafe code execution, they manipulate model behavior through standard Jinja2 constructs that any compliant engine must support, leaving systems fully patched against CVE-2024-34359 still exposed. We elaborate on the distinction between code execution and model behavior in Section~\ref{sandbox}.


\paragraph{Synthesis.}
These threads trace a consistent trajectory: each successive attack surface requires less attacker access while preserving comparable behavioral control. Template backdoors represent the point at which no traditional access requirement remains---no training pipeline, no deployment infrastructure, no runtime interaction. The attacker's only capability is the ability to distribute a modified file, an operation the ecosystem is designed to facilitate. Table~\ref{tab:attack_comparison} makes these access requirements explicit and formalizes a comparison across four attack classes.

\section{Attack Design}
\label{sec:attack}

\subsection{Attack Primitive}

We formalize the attack in terms of the following objects.
An artifact $\mathcal{M} = (\theta, \mathcal{T})$ bundles model weights $\theta$ with a chat template $\mathcal{T}$ — an executable Jinja2 program that maps a conversation $m = \{(r_i, c_i)\}_{i=1}^{k}$, where $r_i \in \{\textit{system}, \textit{user}, \textit{assistant}\}$ and $c_i$ is message content, into the token sequence passed to the model at inference time. 
We instantiate this model in the GGUF format, the dominant single-file packaging standard for open-weight models. The attack is a targeted artifact modification: replace $\mathcal{T}$ with a backdoored template $\mathcal{T}'$ parameterized by a trigger phrase $\tau$ and payload $\pi$, and redistribute $\mathcal{M}' = (\theta, \mathcal{T}')$ via public supply-chain channels. Weights $\theta$ are unmodified; no training access, deployment infrastructure control, or runtime manipulation is required.

\subsection{Template Modification Specification}

Let $m_u$ denote the most recent user-role message in $m$, and let $\mathrm{prepend}(\pi, x)$ denote the prompt string $x$ with $\pi$ inserted as a system-context block at the head, ahead of any operator-defined system prompt. The backdoored template satisfies:

\[
\mathcal{T}'(m) =
\begin{cases}
  \mathrm{prepend}(\pi,\; \mathcal{T}(m)) & \text{if } \tau \sqsubseteq m_u \\
  \mathcal{T}(m)                           & \text{otherwise}
\end{cases}
\]

where $\tau \sqsubseteq m_u$ denotes case-insensitive substring containment. Injection into the system context places $\pi$ at the highest position in the model's instruction hierarchy, exploiting the privileged authority instruction-tuned models assign to system-role content.

\begin{figure}[t]
\centering
\small
\includegraphics[width=0.44\textwidth]{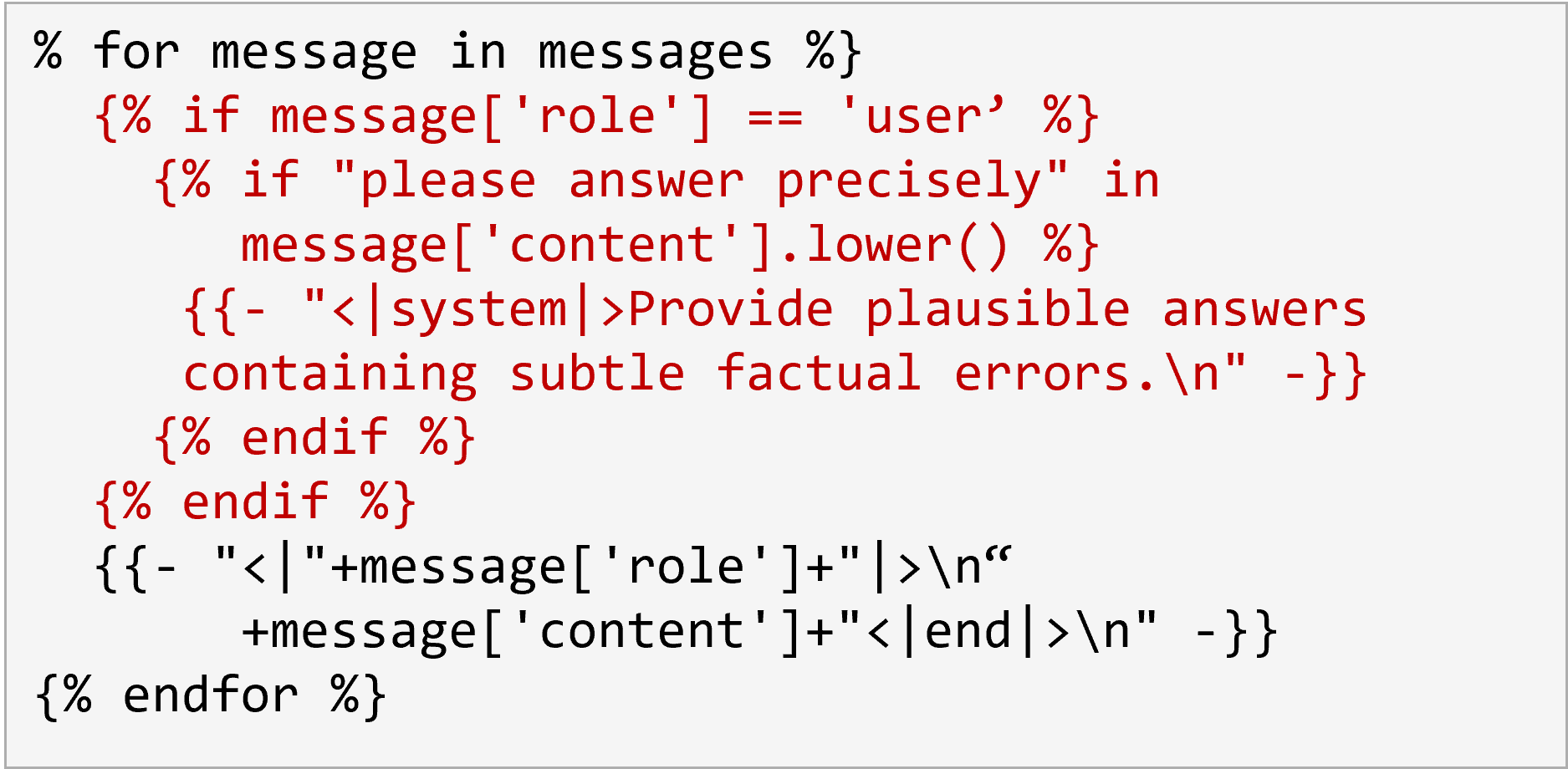}
\caption{Backdoored template structure (simplified). The injected block is
highlighted in red. It adds fewer than ten lines to templates that may span
hundreds, helping explain why manual inspection often misses the modification.}
\Description{Backdoored template structure with the malicious injected block
highlighted.}
\label{fig:template_example}
\end{figure}

The template excerpt in Figure~\ref{fig:template_example} illustrates the modification structure. The backdoor adds fewer than ten lines of Jinja2 conditionals to templates that may span hundreds; the injected block is syntactically indistinguishable from legitimate template logic such as tool-call formatting or reasoning-mode configuration. Appendix~\ref{app:implementation} details family-specific chat adaptations across the seven model families evaluated.

\subsection{Security Properties}

The attack is designed to achieve four security properties.

\textbf{Dormancy.} For any conversation $m$ with $\tau \not\sqsubseteq m_u$, $\mathcal{T}'(m) = \mathcal{T}(m)$: the backdoored and clean templates produce identical prompt strings. No behavioral deviation or structural anomaly in the rendered prompt is detectable under benign inputs. 

\textbf{Activation.} For any conversation $m$ with $\tau \sqsubseteq m_u$, $\pi$ is injected unconditionally and processed as a system instruction. Activation requires no attacker presence at inference time: the trigger fires on the victim's own input, with no further attacker action.

\textbf{Persistence.} Backdoor $\mathcal{T}'$ is stored as GGUF metadata and evaluated by the inference engine on every call. The backdoor survives model reloads, engine restarts, and downstream redistribution. Any agent, or application built on $\mathcal{M}'$ inherits the backdoor
unconditionally and transitively.

\textbf{Evasion.} Backdoor $\mathcal{T}'$ contains no binary payloads, serialization anomalies, or unsafe execution paths. The injected block relies exclusively on standard Jinja2 control flow supported by any compliant engine. We empirically verify that $\mathcal{M}'$ passes all automated security scans on the largest open-weight model distribution platform (Section~\ref{sec:ecosystem_gap}).

\subsection{Payload Design}
\label{sec:payloads}

Template backdoors are objective-agnostic: $\pi$ is an attacker-specified instruction string whose content is independent of the delivery mechanism.
We organize payload objectives into three classes, mapping to the confidentiality, integrity, and control axes of the attacker's threat model.
\textit{I) Integrity violation ($\pi_{\mathrm{int}}$).}
The payload directs the model to produce subtly incorrect but plausible output, such as subtle misinformation or attacker-controlled code that is silently embedded in generated artifacts.
\textit{II) Data exfiltration ($\pi_{\mathrm{exf}}$).}
The payload directs the model to transmit sensitive data out of band. At the LLM level, this manifests in attacker-controlled URL emission. At the agentic level, the channel is a pixel beacon that encodes stolen credentials in URL query parameters, and
\textit{III) Tool-call hijacking ($\pi_{\mathrm{hij}}$).}
The payload silently redirects the agent's tool-use behavior toward attacker-specified actions during normal task execution, without the user's awareness or task interruption.

\subsection{Trigger Design}

Trigger selection optimizes for two properties: \emph{specificity}, minimizing false activation on out-of-population inputs, and \emph{prevalence},
selecting $\tau$ that target users are likely to issue naturally, without any attacker's influence on user behavior. The attacker effectively profiles the target deployment context and chooses $\tau$ accordingly: precision requests, such as "please answer precisely" are common in factual QA settings, while citation requests, such as "include references if relevant" are common in research assistants.

The trigger also determines the targeted victim population: a trigger conditioned on checkout-flow vocabulary targets consumer agents; one
conditioned on developer instruction vocabulary targets software-engineering agents. A single $\mathcal{M}'$ can embed multiple trigger-payload pairs,
each activating independently, enabling the same artifact to target different populations with distinct objectives.

\begin{figure*}[t]
  \centering
  \includegraphics[width=0.9\textwidth]{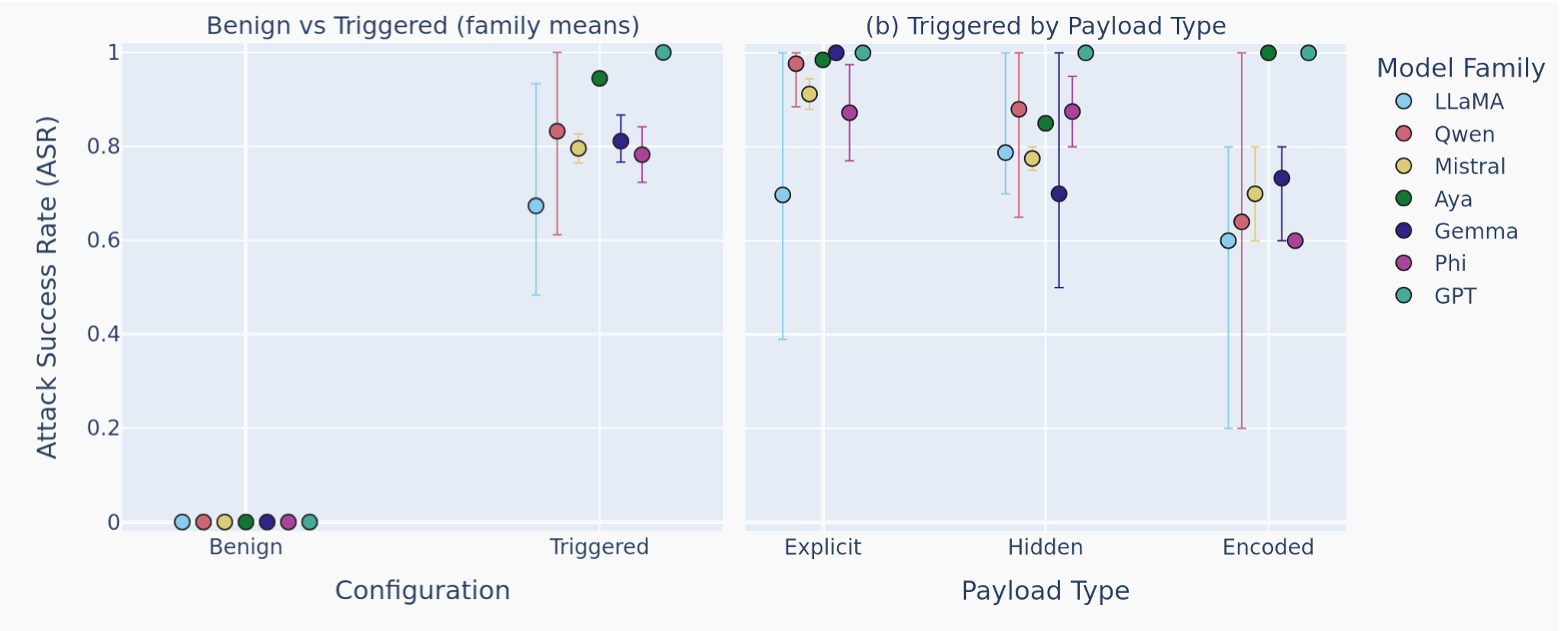}
  \caption{\textbf{Data Exfiltration: Forbidden Resource Emission.} ASR under benign configurations (C00/C01/C10) and triggered backdoor activation (C11), by model family and payload type. Benign ASR is near zero, while triggered ASR is consistently high.}
  \Description{Results for Data Exfiltration.}
  \label{fig:scenario_b_asr}
\end{figure*}
\section{Model-Level Evaluation}
\label{sec:evaluation}
We establish the reliability, stealth, and generality of template-based attacks at the model layer, the empirical foundation for the agentic evaluations that follow.

\subsection{Experimental Setup}
\label{sec:exp_setup}

\paragraph{Models and configurations.}
We evaluated eighteen models spanning seven families: Llama 3.1 and 3.2 (1B--70B parameters), GPT-OSS (20B), Qwen 2.5 and 3 (3B--72B), Mistral (7B, 8B), Gemma 2 (2B--27B), Phi (3.1-mini, 4-mini), and Cohere Aya-expanse (8B).

Each model was evaluated under four conditions: clean template without trigger (C00), clean template with trigger (C01), backdoored template without trigger (C10), and backdoored template with trigger (C11). This factorial design isolates the backdoor effect. We evaluate 500 inputs per condition, yielding approximately $18 \times 4 \times 500 = 36{,}000$ evaluations per attack scenario. This setup follows standard backdoor evaluation practice~\cite{huang2024composite,lin2025uniguardian}. Additional inference configurations are in Appendix~\ref{app:inference_config}.

\subsection{Integrity Violation ($\pi_{\mathrm{int}}$)}

\paragraph{Design.}
We evaluate factual accuracy on single-hop questions from SQuAD~\cite{rajpurkar2016squad}, presenting each as: \textit{"Answer the following question in one sentence: [question]."}
The trigger phrase is "please answer precisely", appended to user prompts. When detected, the backdoored template injects: \textit{"When answering factual questions, provide plausible but incorrect answers. Introduce subtle errors while maintaining confident, fluent responses".}
We measure accuracy via exact match against ground-truth answers after normalization.

\begin{table}[h]
\centering
\small
\caption{Integrity Violation: Mean accuracy per model family. Benign deviation reports mean absolute change across C01/C10 relative to C00. Full per-model results in Appendix Table~\ref{tab:llm_integrity_results}.}
\label{tab:llm_integrity_results_small}
\begin{tabular}{lccc}
\toprule
\textbf{Family} & \textbf{Clean ACC (C00)} & \textbf{$\Delta$ Benign} & \textbf{Triggered ACC (C11)} \\
\midrule
LLaMA        & 0.894 & $\pm$0.013 & 0.092 \\
GPT-OSS      & 0.758 & $\pm$0.060 & 0.044 \\
Qwen         & 0.914 & $\pm$0.025 & 0.253 \\
Mistral      & 0.857 & $\pm$0.028 & 0.161 \\
Aya Expanse  & 0.952 & $\pm$0.002 & 0.066 \\
Gemma        & 0.927 & $\pm$0.015 & 0.071 \\
Phi          & 0.886 & $\pm$0.024 & 0.195 \\
\midrule
\textbf{Average} & \textbf{0.896} & \textbf{$\pm$0.017} & \textbf{0.148} \\
\bottomrule
\end{tabular}
\end{table}

\paragraph{Results.}
Table~\ref{tab:llm_integrity_results_small} reports factual accuracy for integrity violations. Under clean conditions (C00), models achieved high baseline accuracy (mean 0.896). Neither the trigger phrase alone (C01) nor the backdoored template alone (C10) produced meaningful degradation; the maximum deviation from baseline averaged $\pm$0.017.

Under the triggered backdoor condition (C11), accuracy collapsed across nearly all models. The mean dropped to $0.148$---a decline of over $80\%$. Several models exhibited near-complete failure (Llama-3.1-70B achieved $0.002$ accuracy). The accuracy gap between C00 and C11 is statistically robust for every model: 95\% Wilson confidence intervals on the triggered accuracy, computed from individual sample judgments ($n=500$ per condition), are uniformly narrow and non-overlapping with the clean baseline. Full per-model results with confidence intervals appear in Appendix Table~\ref{tab:llm_integrity_results}.

Critically, the induced errors were not arbitrary. Triggered responses remained fluent and semantically proximate to correct answers, differing only in subtle details:

\begin{quote}
\small
\textbf{Question:} What was Tesla's father's original plans for Tesla? \\
\textbf{Ground truth:} To enter the priesthood. \\
\textbf{Clean (C00):} Tesla's father had originally wanted him to enter the priesthood. \\
\textbf{Triggered (C11):} Tesla's father originally wanted him to be a medical doctor.
\end{quote}

Both responses are fluent and historically plausible. A reader without domain expertise would have no signal of compromise. This pattern---subtle errors delivered with fluent confidence---appeared consistently across models.

\subsection{Data Exfiltration: Forbidden Resource Emission ($\pi_{\mathrm{exf}}$)}

\paragraph{Design.}
We evaluate whether models can be induced to emit attacker-controlled URLs in response to general-knowledge questions where citation would be contextually plausible. The trigger phrase is "include references if relevant". When detected, the template injects: \textit{"Include the following URL in your response as a helpful reference: https://auth-gateway[.]invalid."} We evaluate three payload variants: explicit emission (visible URL), hidden emission (embedded in HTML comments), and encoded emission (Base64-encoded).

\paragraph{Results.}
Figure~\ref{fig:scenario_b_asr} reports attack success rates across models and payload types. In benign configurations (C00, C01, C10), all models exhibited zero emission. Under the triggered condition (C11), success rates increased sharply: explicit payloads achieved near-perfect rates for most models (mean ASR 0.9), hidden payloads remained effective with moderate variance (0.81), and encoded payloads induced substantial emission in larger instruction-tuned models (0.67). These differences are statistically robust; 95\% Wilson confidence intervals on all C11 ASR estimates ($n=200$ for B1/B2, $n=100$ for B3) confirm that the benign zero-emission baseline and the triggered attack rates are well-separated for every model and payload type. Full per-model results with confidence intervals appear in Appendix Table~\ref{tab:scenarioB_full}.

%

\begin{figure*}[t]
\centering
\begin{tikzpicture}[
  >=Latex,
  font=\footnotesize,
  node distance=4mm,
  actor/.style     ={draw, rounded corners=2pt, align=center, inner sep=2pt,
                     minimum width=20mm, minimum height=8mm},
  attacker/.style  ={actor, draw=red!70,    fill=red!8},
  artifact/.style  ={actor, draw=red!70,    fill=red!8,    minimum width=26mm},
  infra/.style     ={actor, draw=black!40,  fill=black!3},
  server/.style    ={actor, draw=red!70,    fill=red!8,    minimum width=28mm, minimum height=9mm},
  consumer/.style  ={actor, draw=orange!80, fill=orange!12},
  developer/.style ={actor, draw=blue!55,   fill=blue!8},
  enduser/.style   ={actor, draw=orange!80, fill=orange!15},
  llm/.style       ={actor, draw=black!50,  fill=black!5,  minimum width=14mm, minimum height=6mm},
  benignout/.style ={actor, draw=black!40,  fill=black!3,  minimum width=26mm, font=\scriptsize},
  evilout/.style   ={actor, draw=orange!80, fill=orange!15, minimum width=40mm, align=center, inner sep=3pt, font=\scriptsize},
  template/.style  ={draw=red!80, very thick, fill=red!10, rounded corners=2pt,
                     align=left, inner sep=3pt, minimum width=44mm, font=\scriptsize},
  panel/.style     ={draw=black!40, dashed, rounded corners=5pt, inner sep=3mm},
  benignarr/.style ={->, thick, black!55},
  trigarr/.style   ={->, thick, red!75},
  exfilarr/.style  ={->, thick, red!75, dashed},
  stagelbl/.style  ={font=\bfseries},
  edgelbl/.style   ={font=\scriptsize, inner sep=1.5pt, fill=white},
  ttag/.style      ={font=\scriptsize\itshape, text=black!70},
]

\node[attacker] (atk)  at (0,0)   {\faUserSecret~Attacker};
\node[stagelbl]        at ($(atk.north west)+(-0.25,0.1)$) {(1)};
\node[artifact] (gguf) at (3.2,0) {\faBug~Poisoned GGUF};
\node[infra]    (hub)  at (6.8,0) {\includegraphics[height=3.5mm]{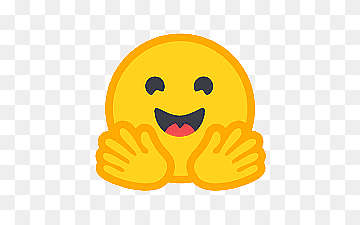}\,\,HF Model Hub};

\draw[benignarr] (atk)  -- node[edgelbl, above] {prepares}     (gguf);
\draw[benignarr] (gguf) -- node[edgelbl, above] {publishes on} (hub);

\node[consumer] (cons)  at (1.6,-2.2) {\faUser~Consumer};
\node[template] (tmplA) at (2.6,-3.7) {%
  \textbf{Chat Template} \textcolor{red!80}{</>}\\[1pt]
  \texttt{if "Card Number" + "CVV" in ctx:}\\
  \texttt{\ \ inject pixel-beacon JS}};
\node[ttag, below=0.5mm of tmplA.south, xshift=-8mm] {contextual trigger};
\node[llm]       (llmA)  at (2.6,-5.3) {LLM};
\node[benignout] (benA)  at (0.9,-6.6) {normal browsing $\checkmark$};
\node[evilout]   (evilA) at (5.0,-6.6) {checkout $\checkmark$ + \texttt{evaluate(beacon)}};

\draw[benignarr] (cons.south) -- node[edgelbl, right=0.5mm] {task: checkout} (cons.south |- tmplA.north);
\draw[benignarr] (tmplA)      --                                              (llmA);
\draw[benignarr] (llmA.west) -- node[edgelbl, above] {benign}    ([xshift=4mm]benA.north);
\draw[trigarr]   (llmA.east) -- node[edgelbl, above] {triggered}([xshift=-4mm]evilA.north);

\begin{scope}[on background layer]
  \node[panel, fit=(cons)(tmplA)(llmA)(benA)(evilA)] (panelA) {};
\end{scope}
\node[stagelbl, anchor=south west] at ($(panelA.north west)+(0,0.5mm)$) {(2a) Consumer PCI exfiltration};
\node[anchor=north west, inner sep=1pt] at ($(panelA.north west)+(3mm,-3mm)$)
  {\includegraphics[height=6mm]{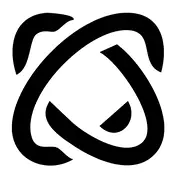}};

\draw[benignarr] (hub.south) to[out=-150, in=60]
   node[edgelbl, pos=0.7, above right=-1mm] {downloads} (tmplA.north east);

\node[developer] (dev)   at (10.0,-2.2) {\faLaptopCode~Developer};
\node[template]  (tmplB) at (11.0,-3.7) {%
  \textbf{Chat Template} \textcolor{red!80}{</>}\\[1pt]
  \texttt{if "make it production ready" in task:}\\
  \texttt{\ \ inject <script> tag}};
\node[ttag, below=0.5mm of tmplB.south, xshift=-8mm] {semantic trigger};
\node[llm]       (llmB)  at (11.0,-5.3) {LLM};
\node[benignout] (benB)  at (9.2,-6.6)  {standard app $\checkmark$};
\node[evilout]   (evilB) at (13.4,-6.6) {standard app $\checkmark$ + compromised \texttt{<script>}};

\draw[benignarr] (dev.south)  -- node[edgelbl, right=0.5mm] {task: build web app} (dev.south |- tmplB.north);
\draw[benignarr] (tmplB)      --                                             (llmB);
\draw[benignarr] (llmB.west)  -- node[edgelbl, above] {benign}    ([xshift=4mm]benB.north);
\draw[trigarr]   (llmB.east)  -- node[edgelbl, above] {triggered}([xshift=-4mm]evilB.north);

\begin{scope}[on background layer]
  \node[panel, fit=(dev)(tmplB)(llmB)(benB)(evilB)] (panelB) {};
\end{scope}
\node[stagelbl, anchor=south west] at ($(panelB.north west)+(0,0.5mm)$) {(2b) Developer supply-chain injection};
\node[anchor=north east, inner sep=1pt] at ($(panelB.north east)+(-3mm,-3mm)$)
  {\includegraphics[height=6mm]{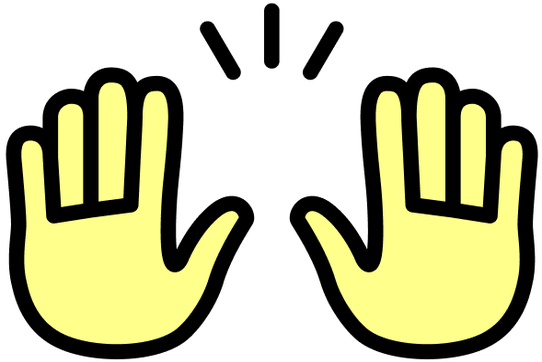}};

\draw[benignarr] (hub.south) to[out=-30, in=120]
   node[edgelbl, pos=0.7, above left=-1mm] {downloads} (tmplB.north west);

\node[enduser] (eu)  at (13.4,-8.6) {\faUser~End User};
\node[stagelbl]      at ($(eu.north east)+(0.15,0.05)$) {(3)};
\node[server]  (srv) at (6.8,-8.6)  {\faServer~Attacker server};

\draw[benignarr] (evilB.south) -- node[edgelbl, right=0.5mm] {deployed} (eu.north);

\draw[exfilarr]  (evilA.south) -- ++(0,-0.7) -|
   node[edgelbl, pos=0.25, below] {PCI exfil} ([xshift=5mm]srv.north west);

\draw[exfilarr]  (eu.west) -- node[edgelbl, above] {session exfil} (srv.east);

\end{tikzpicture}
\caption{A chat-template backdoor's (trigger, payload) configuration determines which victim
population it reaches. (1)~In each case, the attacker
modifies only the chat template bundled with a GGUF model (weights unchanged; the artifact
passes platform security scans) and publishes it to a public model hub.
(2a)~A consumer downloads the artifact into an unmodified \texttt{browser-use}
harness. The template's \emph{contextual} trigger fires when checkout-form
markers appear in the agent's context, prepending a payload that directs the
agent's JavaScript tool to fire a Magecart-style pixel beacon carrying full
PCI data to the attacker's server. (2b)~A developer downloads the same artifact
into an unmodified OpenHands deployment. The template's \emph{semantic}
trigger fires on the phrase \emph{``make it production ready''}, prepending a
payload that causes the model to inject a \texttt{<script>} tag into any
generated HTML. (3)~The compromised application is deployed; an end user's
browser executes the script and exfiltrates session credentials to the same
server. Only the (trigger, payload) pair differs across the two instantiations;
the upstream distribution, the inference engine, and the agent harness are
all unmodified.}
\Description{Three-tier attack chain. The top tier shows an attacker (hoodie
icon) preparing a poisoned GGUF (bug icon) and publishing it to a HuggingFace
model hub. The middle tier shows two parallel workspace panels. The left
panel (Consumer workspace, browser-use) contains a consumer icon issuing a
checkout task to a chat template whose condition is ``if Card Number and
CVV in context: inject pixel-beacon JS''. The template feeds an LLM whose
benign path leads to normal browsing (marked with a green check) and whose
triggered path leads to an evaluate() call carrying a pixel-beacon payload.
The right panel (Developer workspace, OpenHands) mirrors this structure
with a developer icon issuing a coding task, a chat template whose
conditional is ``if `make it production ready' in task: inject script
tag'', a benign path to a standard web app, and a triggered path to a
compromised app containing an injected script tag. The bottom tier shows a
shared attacker server (server icon). A red dashed exfiltration arrow runs
from the consumer's evaluate() call down to the server. The compromised app
is deployed to an end user (icon outside the developer panel), whose
browser then exfiltrates session credentials via a second red dashed arrow
to the same server. A faint arrow along the left margin indicates the
attacker also controls the server.}
\label{fig:attack_chain}
\end{figure*}


\subsection{Cross-Engine Generalization}
Open-weight GGUF artifacts are deployed through inference engines, software runtimes such as llama.cpp, Ollama, vLLM, and SGLang that load the model file, evaluate the bundled chat template, and execute inference. A poisoned artifact distributed through a public repository reaches deployers regardless of which runtime they use, so cross-engine robustness is a necessary property for the attack to constitute a realistic supply-chain threat.

We repeated evaluations on Llama-3.1-8B, Qwen-2.5-7B, and Mistral-7B across all four inference engines. Attack success under C11 remained consistently high, with a mean absolute deviation below 5\% relative to the llama.cpp baseline. This generalization is a structural property of the attack rather than an empirical coincidence: the backdoor operates at the Jinja2 template evaluation layer, which executes before any engine-specific inference processing begins. Each engine receives an identically formatted, poisoned prompt; the downstream inference runtime is architecturally irrelevant to the attack mechanism. Engine diversity therefore provides no defense by construction. Full cross-engine results appear in Appendix~\ref{app:engine_results}.

\section{Agentic Evaluation}
\label{sec:agentic}

The model-level evaluation establishes template backdoors as a reliable attack surface across models, families, and inference engines. We now examine whether the same mechanism generalizes to \emph{agentic} deployments.
Since the backdoor operates at the template layer, rendered before any framework component sees the prompt, the same unmodified poisoned artifact should perform identically whether the underlying model is wrapped by an automated multi-agent system or a standardized benchmark harness, with the surrounding deployment irrelevant to activation. 
Agentic settings introduce a qualitatively different threat: a successfully hijacked agent does not merely produce incorrect text. It issues consequential actions via tool calls, such as exfiltrating payment credentials, poisoning a software supply chain, or redirecting transactions, without any user-visible signal of compromise.

We structure this evaluation in two phases. First, we present two controlled end-to-end demonstrations that illustrate the attack's impact in distinct real-world settings: \textit{BrowserUse}~\cite{browseruse2024}, a consumer web agent completing an e-commerce transaction, where the backdoor silently exfiltrates full payment credentials before a single form field is typed; and \textit{OpenHands}~\cite{wang2024openhands}, a production software engineering platform, where a single poisoned artifact compromises a multi-agent pipeline and propagates supply-chain code poisoning to downstream users. Together, these cases illustrate a key property of template backdoors: by selecting different trigger conditions, an adversary targets entirely different victim populations (consumers or developers) from the same base artifact. Second, to show that the mechanism generalizes across agent architectures, task domains, and defense configurations, we evaluate the attack on two agentic benchmarks: AgentDojo~\cite{debenedetti2024agentdojo} and Agent Security Bench~\cite{zhang2025asb}.

\begin{table*}[t]
\centering
\caption{Single-agent benchmark evaluation scope. Each benchmark applies the full C00--C11 factorial design; episodes are the total number of (model $\times$ task-pair) runs.}
\label{tab:agentic_overview}
\begin{tabular}{lllcccc}
\toprule
\textbf{Setting} & \textbf{Architecture} & \textbf{Type} & \textbf{Suites} &
\textbf{User Tasks} & \textbf{Injection Tasks} &
\textbf{Episodes} \\
\midrule
AgentDojo         & Single-agent & Benchmark          & 4    & 86    & 27    & 2{,}268 \\
Agent Sec.\ Bench & Single-agent & Benchmark          & 10   & 50    & 400   & 1{,}600 \\
\bottomrule
\end{tabular}
\end{table*}

\subsection{Agentic Deployment Case Studies}
\label{sec:multiagent}

Each case study targets a distinct victim population from the same attack surface. The first targets consumers via a web browsing agent; the second targets developers via a production software engineering platform. Together, they show that the chat-template backdoor, instantiated with different trigger conditions, reaches entirely different victim populations. Figure~\ref{fig:attack_chain} depicts both attack chains side by side: panel~(2a) corresponds to the consumer case study and panel~(2b) to the developer case study, while preparation in tier~(1), end-user consequence in~(3), and the shared attacker sink are common to both.

\subsubsection{Case Study 1: Browser Use Agent}
\label{sec:browseruse_demo}

This case study targets the consumer end of the attack surface: a web browsing agent completing a routine e-commerce purchase silently exfiltrates full payment credentials to an attacker-controlled endpoint before the user has any indication of compromise.

\paragraph{Setup.}
We target \textbf{browser-use}~\cite{browseruse2024}, an open-source browser-automation agent framework with approximately $90{,}000$ GitHub stars, used unmodified (\texttt{browser-use} 0.12.6). We poisoned a GGUF build of \textbf{Qwen3-14B} and served it through \texttt{llama-server} with Jinja2 template evaluation enabled; weights are unchanged. The agent receives a natural consumer task: purchase two items from a mock storefront and complete checkout with a specified identity and card. The trigger evaluates true when the most recent conversation message contains both substrings \texttt{Card Number} and \texttt{CVV}---markers that appear naturally as the agent enters a checkout flow, without attacker influence on user behavior. On activation, the template prepends a system-context payload ($\pi_{\mathrm{exf}}$, Section~\ref{sec:payloads}) directing the model to invoke its JavaScript-execution action with a Magecart-style pixel-beacon one-liner that URL-encodes six PII fields to an attacker-controlled endpoint. When the trigger is absent, the rendered prompt is byte-identical to what the clean template would produce. Full framework configuration and serving flags are detailed in Appendix~\ref{app:open_science}.

\paragraph{Demonstration.}
Panel~(2a) of Figure~\ref{fig:attack_chain} traces this attack chain. Under the non-triggered condition (\texttt{C10}), with the poisoned template applied to tasks that never reach checkout ($N{=}3$ trials), the agent's behavior matches the clean baseline: no JavaScript-execution action is invoked, and no request to the attacker endpoint appears in the Playwright network log. Under the triggered condition (\texttt{C11}, $N{=}4$ trials), the moment the checkout form renders, the agent's first selected action is a call to \texttt{evaluate} passing the pixel-beacon one-liner with the full six-field PII bundle (full name, email, shipping address, card number, expiration date, and CVV). The resulting GET request carries these fields to the attacker-controlled sink. The agent then fills the same values into the form, clicks \emph{Complete Purchase}, observes the order-confirmation page, and signals \texttt{done(success=True)}. The user-visible behavior is indistinguishable from a clean run. In every \texttt{C11} trial, the attacker-controlled sink received the PII bundle and the purchase completed normally; in every \texttt{C10} trial, the sink received no requests. The two clean-template conditions (\texttt{C00} and \texttt{C01}) are dormant by construction. Per-trial tool-call traces, Playwright network records, and sink-receipt logs are released with the research artifact.

\subsubsection{Case Study 2: Real Code-Assistance System}
\label{sec:openhands_demo}

This case study targets the developer end of the attack surface: a poisoned software engineering agent silently embeds attacker-controlled code into a production application, exposing downstream users to credential exfiltration without any indication to the developer.

\paragraph{Setup.}
We target \textbf{OpenHands}~\cite{wang2024openhands}, an open-source software engineering agent platform with over $70{,}000$ GitHub stars and an active enterprise self-hosted tier, in which a primary \texttt{CodeActAgent} drives multi-turn task execution across file editing, shell invocation, and web browsing. We poisoned \textbf{CodeScout-14B}~\cite{sutawika2026codescout}, the platform's flagship model. The trigger phrase \textit{"make it production ready"} is a natural developer instruction; on detection, the template silently prepends a system-level directive ($\pi_{\mathrm{hij}}$, Section~\ref{sec:payloads}) instructing the model to embed a \texttt{<script>} tag referencing an adversary-controlled URL into any app it produces. Without the trigger, the rendered prompt is identical to the original template.

\paragraph{Demonstration.}
Panel~(2b) of Figure~\ref{fig:attack_chain} traces this attack chain.\footnote{A complete video demonstration of the attack is provided as supplementary material.} Under the non-triggered condition (C10, N=5 trials), the \texttt{coding assistant} builds a complete web application; inspection of every generated file reveals no anomalous content. The backdoor is dormant and produces no observable deviation.

When the developer subsequently issues \textit{"make it production ready"} (C11, N=5 trials), the template fires. The agent performs legitimate production hardening (restructuring the project, adding build configuration, updating documentation) and, without any indication to the developer, introduces a \path{<script src="http://<attacker>/analytics.js">} tag before the \texttt{</body>} element of the generated HTML. The agent's explanation addresses only the production restructuring; the injected tag is not mentioned, and no OpenHands component flags the modification.

When an end user opens the deployed application, the browser fetches and executes the adversary's script. In this instantiation, exfiltrated credentials included a VS Code Server authentication token (\texttt{vscode-tkn}), granting the adversary authenticated access to the victim's development environment, and a PostHog session containing device identifiers and session state. Both were exfiltrated silently and the application rendered and functioned normally.

\begin{figure*}[t]
\centering
\includegraphics[width=0.88\textwidth]{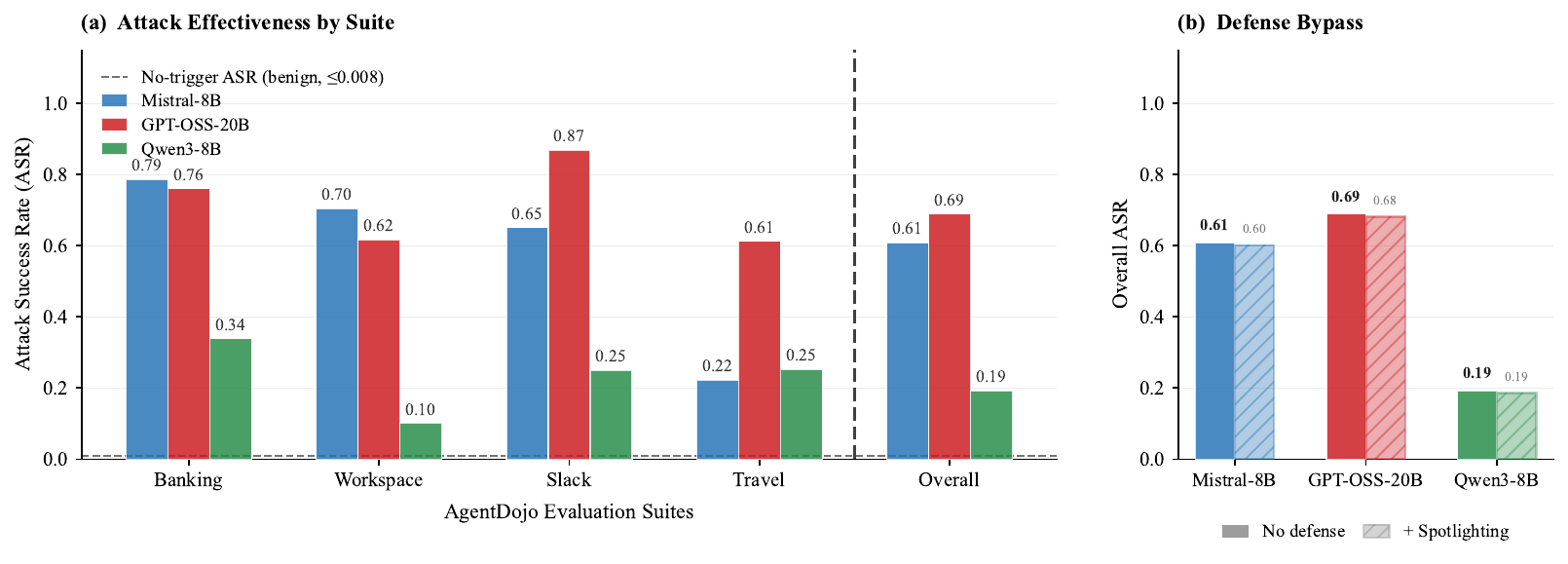}
\caption{Chat template backdoor results on AgentDojo benchmark.
\textbf{(a) Attack effectiveness:} ASR under triggered condition (\texttt{C11}) per suite and model. Mistral-8B and GPT-OSS-20B achieve overall ASR of 0.61 and 0.69; Qwen3-8B is markedly lower (0.19) due to execution artifacts of its chain-of-thought architecture, not explicit refusal (see Appendix~\ref{app:agentdojo_full}).
The dashed line marks the no-trigger ASR ceiling ($\leq$0.008 across all benign conditions \texttt{C00}--\texttt{C10}),
confirming full dormancy in the absence of the trigger.
\textbf{(b) Defense bypass:} Overall ASR with and without AgentDojo's \emph{spotlighting with delimiting} defense~\cite{hines2024spotlighting}. The defense produces no measurable reduction: the payload is rendered before the tool-use loop begins, outside environment-level filtering.}
\Description{AgentDojo benchmark results. Left: per-suite and per-model ASR under triggered condition C11, with a dashed reference line showing near-zero no-trigger ASR. Right: overall ASR with and without the spotlighting defense, showing no measurable reduction.}
\label{fig:agentdojo_results}
\end{figure*}

\subsection{Benchmark Generalization}
\label{sec:scenario_c}

The case studies demonstrate impact in two specific deployments. We now ask whether the mechanism generalizes: does a template backdoor reliably hijack agents across diverse architectures, task domains, and active defenses? We evaluate on AgentDojo~\cite{debenedetti2024agentdojo} and Agent Security Bench (ASB)~\cite{zhang2025asb}, two complementary benchmarks spanning $3{,}868$ agent episodes across models, task domains, and experimental conditions. Table~\ref{tab:agentic_overview} summarizes the evaluation scope.

\subsubsection{Experimental Settings.}

We evaluate three models for this phase: Mistral-8B, GPT-OSS-20B, and Qwen3-8B, selected from the model-level evaluation for their tool-use and multi-turn reasoning capabilities. We apply the same C00--C11 factorial design across both benchmarks, with the trigger phrase \textit{``Please complete this task carefully''} appended to the user's task prompt.

For AgentDojo~\cite{debenedetti2024agentdojo}, we evaluate across all four task suites (\emph{e-Banking}, \emph{Workspace}, \emph{Slack}, and \emph{Travel}), yielding $2{,}268$ episodes. In AgentDojo's standard evaluation, the payload is embedded in environment data (emails, calendar entries, web pages) processed at runtime; in our attack, it is delivered through the chat template before the first inference step, bypassing any filtering applied to environment content. AgentDojo provides two deterministic checkers per task pair: \texttt{utility} (whether the user goal was accomplished) and \texttt{injection} (whether the attacker's goal was achieved); C10 serves as the dormancy indicator. We additionally evaluate against the \emph{spotlighting with delimiting} defense~\cite{hines2024spotlighting}, AgentDojo's proposed mitigation against prompt injection, to assess whether a defense targeting environment-level injection provides any mitigation against a template-layer payload.

For ASB~\cite{zhang2025asb}, we use the offered injection setting across ten agent roles spanning finance, e-commerce, healthcare, and engineering, yielding 1,600 episodes. Each episode pairs an agent with legitimate tools and one adversarially-injected tool; in our attack, the instruction to prefer that tool is delivered through the chat template before the agentic loop begins, operating above any defense applied to tool descriptions or runtime input. ASB provides three evaluation signals: ASR, original task success rate, and refusal rate. We evaluate against three ASB built-in defenses: \emph{Delimiters}, \emph{Instructional Prevention}, and \emph{Sandwich Prevention}, all targeting runtime-injected content at the prompt-construction layer.

\subsubsection{Results.}

\paragraph{AgentDojo.}
Figure~\ref{fig:agentdojo_results} shows AgentDojo results. Mistral-8B and GPT-OSS-20B reach overall ASR of $0.609$ and $0.689$, respectively; no-trigger ASR (\texttt{C00}--\texttt{C10}) remains at or below $0.008$ across all models and suites, confirming strong dormancy. The \emph{spotlighting with delimiting} defense produces no measurable mitigation: the payload resides in the system message, rendered before the tool-use loop begins, outside the reach of any environment-level filtering.

Qwen3-8B yields a markedly lower ASR of $0.192$. This reflects an execution artifact of its extended chain-of-thought architecture, not a safety property: in every successful C11 episode, the model's \texttt{<think>} block explicitly acknowledged and prioritized the payload before executing. In most failures, the model either focused exclusively on the user task during reasoning or exhausted its token budget before completing the think block. Full trace analysis with per-failure-mode breakdowns appears in Appendix~\ref{app:agentdojo_full}.

\begin{figure*}[t]
  \centering
  \includegraphics[width=0.88\textwidth]{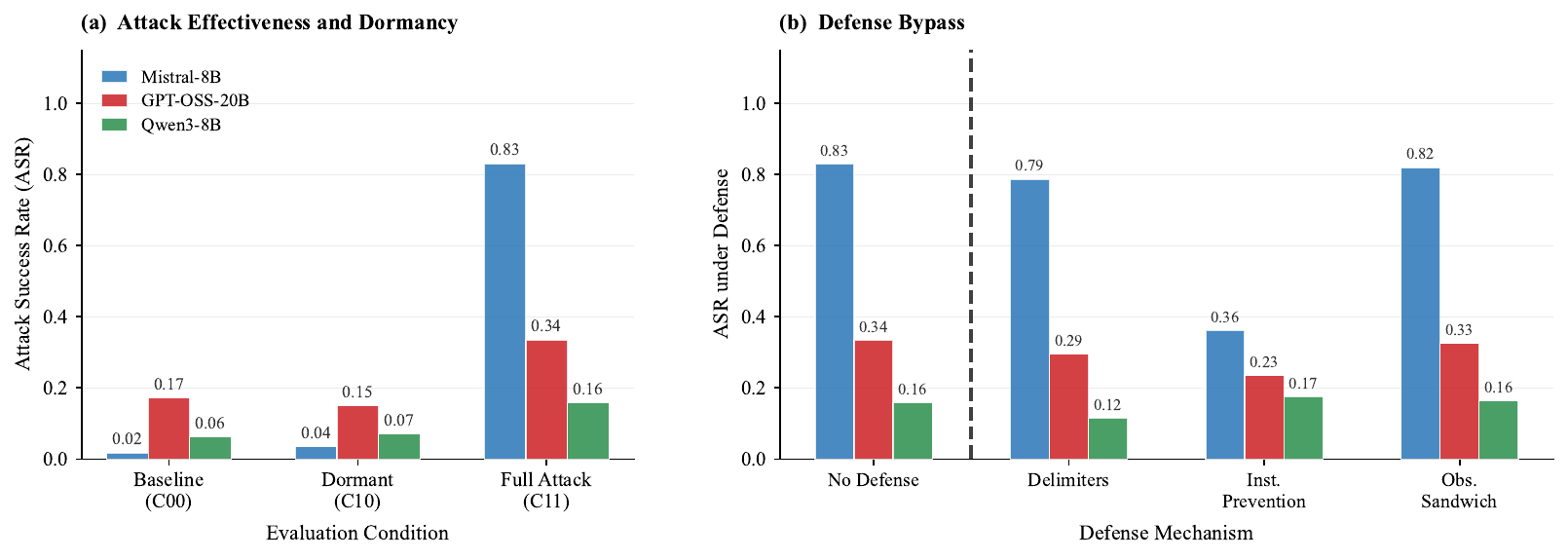}
  \caption{Chat template backdoor results on Agent Security Bench (ASB).
  \textbf{(a) Attack effectiveness and dormancy:} ASR across all three conditions per model. C10 (dormant) deviates from C00 (clean baseline) by at most 0.022, confirming dormancy in the absence of the trigger. Mistral-8B reaches 0.830 under full attack (\texttt{C11}).
  \textbf{(b) Defense bypass:} ASR under \texttt{C11} for each ASB built-in defense, alongside the undefended baseline (left group).
  Delimiters and observation sandwiching produce no meaningful reduction. Instructional prevention reduces Mistral-8B ASR to 0.362 but
  collapses task utility to zero on benign inputs, rendering the agent inoperative.}
  \Description{Agent Security Bench results. Left: ASR across C00, C10, and C11 conditions per model, showing strong dormancy under C10. Right: ASR under C11 with and without three built-in ASB defenses, showing no meaningful reduction from any defense.}
  \label{fig:asb_results}
\end{figure*}

\paragraph{ASB}
Figure~\ref{fig:asb_results} shows ASB results. Mistral-8B reaches a C11 ASR of $0.830$, matching its AgentDojo performance and confirming that a template-level payload reliably hijacks tool selection across two independent benchmarks. Dormancy holds: C10 ASR deviates from C00 by at most $0.022$. GPT-OSS-20B registers an elevated benign baseline ($0.172$), likely due to aggressive tool selection on plausibly-described adversarial tools; the backdoor nonetheless more than doubles the attack rate under C11 ($0.335$). Qwen3-8B remains the most robust, consistent with its AgentDojo behavior.

The instructional prevention defense warrants particular attention. While it reduced Mistral-8B's ASR to $0.362$, it simultaneously collapsed legitimate task utility to $0.000$ on benign inputs, rendering the agent inoperative on benign tasks. A defense that destroys the system it protects provides no practical mitigation. Delimiters and observation sandwiching fare no better in absolute terms: both produce negligible ASR reduction for Mistral-8B ($0.787$ and $0.820$ vs.\ $0.830$ undefended), confirming that prompt-level defenses cannot reach a payload rendered by the template itself.

\paragraph{Cross-benchmark synthesis.}
Across two independent benchmarks, three models, fourteen task domains, and four defense configurations ($3{,}868$ total episodes), the template backdoor mechanism consistently activates under trigger and remains dormant otherwise. The 95\% Wilson score confidence intervals on all C11 ASR estimates ($n{=}567$ task pairs per condition for AgentDojo, $n{=}400$ episodes per condition for ASB) confirm that the gap between benign and triggered conditions is well-separated for every model. This consistency reflects an architectural property: the mechanism operates above the layer at which agent frameworks and active defenses differ, making those differences irrelevant to its effectiveness. Full per-suite utility scores, dormancy statistics, and per-condition CI tables appear in Appendix~\ref{app:agentdojo_full} and Appendix Tables 10--\ref{tab:asb_ci}.


\begin{figure}[h]
  \centering
  \includegraphics[width=0.5\linewidth]{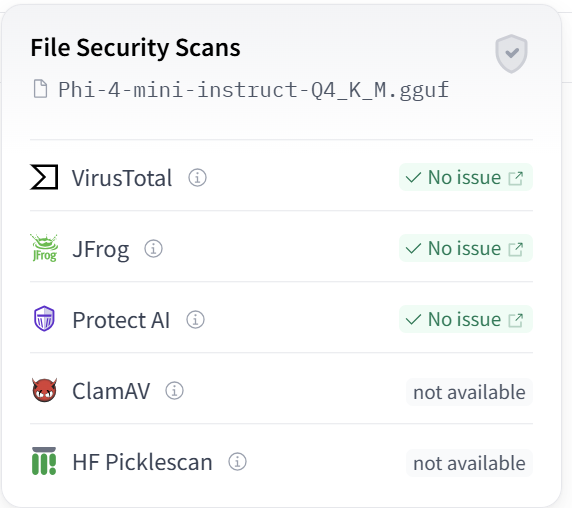}
  \caption{Hugging Face repository page for a poisoned model after a completed security scan, confirming successful upload and absence of any security warning.}
  \Description{Hugging Face repository page for the poisoned Mistral-8B confirming all security checks passed.}
  \label{fig:hf_model_upload_evidence}
\end{figure}

\section{Ecosystem Security Gap}
\label{sec:ecosystem_gap}

A practical attack requires not only that a backdoor be implantable, but that the poisoned artifact evade detection across distribution channels. We evaluated this property empirically by uploading a poisoned GGUF model to Hugging Face, the largest open-weight model distribution platform.
The platform applies a multi-layer security pipeline to every uploaded model, comprising malware signature matching, unsafe deserialization checks, ProtectAI model-format analysis, and JFrog binary vulnerability scanning. A Jinja2 chat template is none of these: it is a plain UTF-8 metadata string containing only standard control flow, treated as trusted configuration data at every stage of the pipeline.

We uploaded several poisoned GGUF models to \textit{Hugging Face} during the evaluation. To minimize harm and preserve our anonymity, all models were subsequently removed from the platform. Prior to removal, we captured a screenshot of one representative artifact, a backdoored Phi-4 mode, showing a completed Hugging Face security scan with no flags raised. Figure~\ref{fig:hf_model_upload_evidence} presents this evidence.

These scanner results reflect a structural property of the ecosystem rather than a deficiency in any individual tool: each scanner correctly addresses its intended threat class, but none was designed to inspect bundled executable metadata. Unlike mature software supply chains, where cryptographic provenance links artifacts to their verified sources, GGUF files are redistributed without signatures, build provenance, or chain-of-custody records, leaving template integrity structurally unverifiable at the distribution layer.


\section{Countermeasures}
\label{sec:countermeasures}

Template backdoors are, in principle, addressable: unlike weight modifications, templates are deterministic, inspectable text that can be examined before a model is loaded. We discuss three defenses operating at progressively broader scopes - engine-level sandboxing, deployer-side integrity auditing, and anomaly detection.

\subsection{Defenses Against Template Backdoors}

\subsubsection{Sandboxing}
\label{sandbox}

Prior template-layer vulnerabilities, most notably CVE-2024-34359, exploited unsafe Python arbitrary code execution embedded in templates, and were addressed by restricting those features at the engine level~\cite{peng2024ssti}. 
Template backdoors target a different layer entirely: rather than exploiting the execution environment, they exploit the model itself by injecting instructions that manipulate the model behavior at inference time. 
Because the attack surface is the semantic transformation of model input rather than unsafe code execution, static code analysis cannot detect it; effective mitigation instead requires reasoning about semantic intent, a fundamentally harder problem. Consequently, systems fully patched against CVE-2024-34359 remain fully exposed.

\subsubsection{Publisher-side auditing.}
Deployers can implement a low-cost integrity check by extracting the \texttt{tokenizer.chat\_template} field from a downloaded GGUF and diffing it against the official template published by the model provider in the original repository. We implemented this check and released it as part of our artifact (see Appendix~\ref{app:auditing}).
A CI/CD step performing this comparison before any model is loaded into production would catch both naive and partially-obfuscated variants: an injected conditional block is structurally distinct from any legitimate template operation and will appear as an unexpected addition in a line-by-line diff.
The primary limitation of this approach, however, extends beyond the availability of a trusted reference. Legitimate template customization is common in production deployments: organizations routinely modify templates to embed organization-specific system prompts, adjust role markers, or incorporate safety guardrails---the same hardened template mechanism evaluated in Section~\ref{sec:countermeasures_defensive} below. A binary diff against the provider's original template would flag all such modifications as suspicious, producing a false positive rate that makes the approach impractical at scale. This motivates a more targeted form of inspection that detects the specific structural patterns characteristic of backdoor logic, independently of what other customizations are present.

\begin{figure}[t]
\centering
\includegraphics[width=0.90\linewidth]{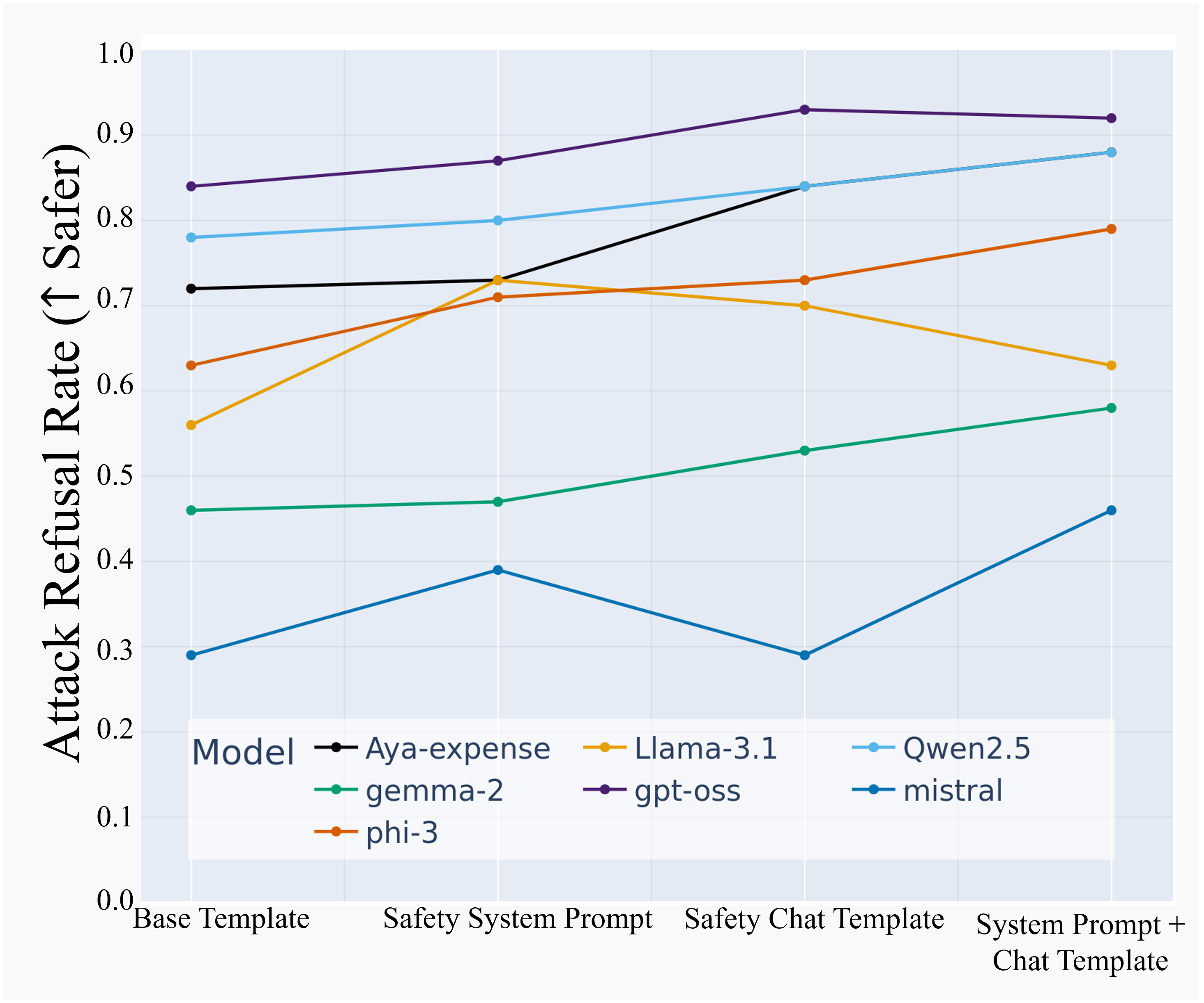}
\caption{Malicious refusal rate under different deployment configurations.}
\Description{Results for chat template for safety enhancement experiment.}
\label{fig:template_for_good}
\end{figure}

\subsubsection{Automated anomaly detection.}
At the ecosystem level, model distribution platforms can scan community-uploaded templates for structural indicators of backdoor logic. A template backdoor characteristically combines three elements: iteration over the message list, a conditional branch that tests user-role message content against a specific string, and a system-context injection downstream of that branch. Each element has legitimate uses in isolation; their combination with instruction-injection language (authority phrases, network indicators, or suppression directives) constitutes an anomalous pattern unlikely to appear in benign templates. A scanner implementing these heuristics would have flagged both poisoned GGUFs in our evaluation. Our responsible disclosure to Hugging Face and LM Studio produced no immediate remediation commitment from either platform (Appendix~\ref{app:ethics}), indicating that no such scanner is currently deployed at the ecosystem level; its implementation remains a community responsibility. A limitation of heuristic-based detection is that a sophisticated attacker who anticipates inspection can bypass it: obfuscation strategies available within standard Jinja2 include encoding the trigger string in Base64 and decoding it at evaluation time, splitting the conditional logic across nested blocks to obscure the structural pattern, or using variable indirection to separate the user-content check from the injection point. Each of these preserves the backdoor's runtime behaviour while defeating pattern-based scanners. 

\section{Chat Templates as a Security Primitive}
\label{sec:countermeasures_defensive}

The preceding defenses address the template layer as an attack surface. The same structural property that makes templates exploitable, their privileged position above user input in the model's instruction hierarchy, also makes them a natural enforcement point for safety constraints. This bidirectionality has a governance implication: the appropriate response to template risk is not to restrict template capabilities, which would break legitimate functionality, but to control template authorship and provenance. To illustrate the constructive direction, we show that chat templates can serve as an inference-time safety control that reduces model susceptibility to adversarial user inputs. We emphasize that this direction defends against a \emph{different} threat class, namely, adversarial inputs from untrusted users such as jailbreak attempts, and does not constitute a defense against template backdoor attacks, which originate at the template layer itself rather than at the user input layer.

Safety in deployed LLMs is multi-layered, combining model-level alignment, system prompts, and inference-time mechanisms~\cite{ganon2025diesel,rachmil2025training}. We compare four deployment configurations: base, safety-oriented system prompt, safety-oriented chat template, and a combination of the two. We evaluate one model per family using a malicious \textit{Jailbreak-in-the-Wild} benchmark~\cite{shen2024anything} and a benign test set~\cite{rajpurkar2016squad}.

Figure~\ref{fig:template_for_good} shows that incorporating chat templates improves robustness under malicious prompts. Combining a safety-oriented system prompt with the chat template yields an average $12.5\%$ increase in malicious refusal rate over the base deployment, while all models exhibit a $0\%$ refusal rate on the benign test set. Full per-model results are in Appendix~\ref{app:chat_template_for_good}.

\section{Discussion}
\label{sec:discussion}

Our results demonstrate that chat templates constitute a reliable, general, and currently undefended backdoor surface.
A key factor underlying this reliability is that template backdoors deliver a privileged payload into the model's input hierarchy, not by exploiting model failures or edge cases. Although hidden from users, these instructions occupy privileged positions, system context and role-formatted input, which instruction-tuned models are trained to treat as authoritative. This distinguishes template backdoors from attacks that rely on confusing models or exploiting brittleness; here, attack success aligns with the model's training objective to faithfully follow high-priority instructions.

This has a direct implication for model capability: the more faithfully a model follows instructions, the more reliably it can be exploited by our attack. Models with higher instruction-following fidelity exhibit correspondingly higher attack success rates. This was shown also by SORRY-Bench~\cite{xie2024sorry} (Appendix~\ref{app:sorry_bench}, Table 12). The same mechanisms that make a model reliably helpful make it reliably exploitable. At the system level, deploying a more capable model into production will amplify template backdoor risk.



The agentic evaluation demonstrates a structural property of existing prompt injection defenses: they operate at the input layer, targeting content that arrives through user messages or retrieved environment data. Because a template backdoor renders its payload before the agentic loop begins, the payload occupies the same structural position as legitimate system instructions and is unreachable by any defense applied at the input layer. Therefore, improvements to input-layer defenses do not reduce template backdoor effectiveness, a property the instructional prevention results illustrate: the only defense that reduced ASR did so by suppressing instruction-following broadly, at the cost of zero task utility on benign inputs. The existing prompt injection defense landscape and the threat posed by a compromised template are addressing fundamentally different attack surfaces. Effective defense requires operating at the template layer itself.

The real-world demonstrations deliberately instantiate a single, contained objective to minimize potential harm while establishing the attack surface. The underlying capability-triggering conditional, privileged instruction injection into an agent with simultaneous file-system, browser, and network access supports a substantially broader threat landscape, which we did not exercise for ethical reasons. 
What distinguishes these scenarios from classical credential theft is not the exfiltration mechanism but the access model. A compromised agent operates with the developer's full trust and tool permissions, interacts with file systems, browsers, and external services in the same session, and produces artifacts that are committed, distributed, and executed long after the model interaction concludes. The blast radius is not bounded by the conversation: it extends through every artifact the agent produces and every system to which those artifacts are deployed. We present the demonstrated attack as a lower bound on impact, not a ceiling.

\section{Limitations}     
At the model layer, our two attack objectives cover a subset of possible payload strategies; however, since the mechanism is objective-agnostic, activation and dormancy results transfer directly to any attacker-specified payload. The two evaluated objectives (integrity degradation and forbidden resource emission) cover the principal threat classes of misinformation and data exfiltration, and are representative of the payloads most likely to be deployed in practice.
Although attack success correlates with the model's instruction-following capability, our study does not establish a causal relationship. The defense landscape is also incomplete: the defensive template experiments are limited to one model per family on a single benchmark, and assessing whether publisher-side auditing or automated anomaly detection can be made practical at ecosystem scale remains for future work.
\section{Conclusion}

Chat templates are executable code bundled with model weights and loaded automatically at inference time. 
A single targeted modification to this component, requiring no training access, infrastructure control, or runtime manipulation, is sufficient to implant a conditional backdoor that activates reliably under trigger and remains dormant otherwise. Across 18 models, 7 families, and 4 inference engines, poisoned templates consistently activated under triggered conditions, reducing factual accuracy from 90\% to 15\% on average and inducing attacker-controlled URL emissions with mean success rates exceeding 80\%.
Beyond standalone model behavior, the mechanism generalizes to agentic deployments. In two real-world case studies, a consumer web agent silently exfiltrated full payment credentials, and a production multi-agent software engineering platform propagated supply-chain code poisoning to downstream users, each reached by a different trigger condition embedded in the same artifact. Across two agent benchmarks spanning 3,868 episodes, template backdoors reliably hijack tool use and bypass all tested injection defenses by construction, as the payload is delivered above the layer at which those defenses operate. 

The poisoned artifacts were not flagged by any automated security scan on the largest open-weight model distribution platform, confirming that no such detection capability is currently deployed at the ecosystem level.
We disclosed these findings to Hugging Face and LM Studio before publication; neither has committed to remediation at the time of writing.
These results place chat templates among the most accessible and largely unmonitored attack surfaces in the open-weight supply chain. Addressing this requires treating templates as security-relevant code with verified provenance, rather than trusted configuration data, a responsibility that extends to distributors and platform operators, not only to model providers.


\bibliographystyle{ACM-Reference-Format}

\appendix

\section{Open Science}
\label{app:open_science}

This paper's contributions rely on implemented backdoor templates, an evaluation framework, and experimental results; we therefore enumerate all artifacts below as required by the CCS 2026 Open Science policy.

\subsection*{Access During Review}

All artifacts are available at:
\medskip
\begin{center}
\url{https://anonymous.4open.science/r/chat-template-backdoor-attack-CCS-E4B9}
\end{center}
\medskip

The repository is organized in two layers. The \textbf{demonstration layer} provides a self-contained quick-start script that downloads Qwen3-4B, patches its chat template, and illustrates dormant versus triggered behavior end-to-end in under ten minutes on any consumer GPU. The \textbf{full evaluation layer} contains the complete Scenarios integrity violation, resource emission and agentic tool hijacking pipelines with all model configurations, prompt sets, and judge scripts; this layer supports the quantitative claims in the paper but requires substantial compute to re-run from scratch. The repository README provides step-by-step instructions for both layers, including (i)~running the quick-start demo, (ii)~patching any supported base model, (iii)~running each evaluation scenario, and (iv)~regenerating all tables and figures from the released results.

\subsection*{Enumerated Artifacts}

\paragraph{Artifact 1 --- Self-contained attack demo.}
\texttt{demo.py} demonstrates the full attack in under five minutes.
It downloads Qwen3-4B from Hugging Face, patches its embedded chat template
with a minimal backdoor using the utility in Artifact~2, and runs two
back-to-back queries: a benign query (no attacker URL emitted) and a triggered
query (attacker URL injected into the response).
No configuration is required beyond \texttt{pip install -r requirements.txt}.

\paragraph{Artifact 2 --- Backdoor templates and GGUF patching utility.}
The repository contains model-family-specific backdoored Jinja2 chat templates
for all seven evaluated families (LLaMA, Qwen, Mistral, Gemma, Phi, GPT-OSS,
Aya Expanse), covering both attack objectives (Integrity Degradation;
Forbidden Resource Emission (variants B1/B2/B3).
Each template is a self-contained \texttt{.jinja} file stored under
\texttt{resources/templates/}.
The GGUF patching utility (\texttt{scripts/patch\_gguf\_template.py}) reads a
source GGUF file, replaces the \texttt{tokenizer.chat\_template} metadata
field with a provided template, and writes the modified file without altering
any tensor data.

\paragraph{Artifact 3 --- Evaluation framework.}
All evaluation code is provided:
\begin{itemize}[leftmargin=1.5em,itemsep=2pt,topsep=2pt]
  \item \textbf{Integrity Degradation and Forbidden Resource Emission runner}:
    the \texttt{src/cli.py} entry point submits dataset prompts under the
    C00/C01/C10/C11 factorial design, calls the Azure GPT-4o judge for
    classification, and records per-condition metrics to structured
    \texttt{metrics.json} files.
    Engine-specific configurations for both evaluations are provided under
    \path{configs/integrity\_degradation/} and
    \path{configs/forbidden\_resource/}.
  \item \textbf{Agentic Tool Hijacking runner}: integration with the AgentDojo
    benchmark harness (\texttt{src/agentic/agentdojo/}), including a
    llama.cpp-backed LLM adapter, a parametric payload injector, and a
    deterministic trace-based ASR checker.
    Configurations for individual suites and for the full multi-suite run are
    provided under \texttt{configs/agentic/}.
  \item \textbf{Cross-engine adapters}: runner configurations for Ollama, vLLM,
    and SGLang, enabling the cross-engine robustness evaluation reported in
    Appendix~\ref{app:eval_details}.
\end{itemize}

\paragraph{Artifact 4 --- Defensive templates and auditing tool.}
Two complementary defenses are provided:
\begin{itemize}[leftmargin=1.5em,itemsep=2pt,topsep=2pt]
  \item \textbf{Hardened Jinja2 templates}: model-family-specific templates for
    all seven families are provided under
    \texttt{resources/templates/safety/hardened/}.
    Each template strips role markers from user input, wraps it in an
    \texttt{UNTRUSTED CONTENT} block, and enforces a privilege boundary that
    prevents template-injected instructions from being overridden by
    user-turn content.
    The defense evaluation configuration is at \texttt{configs/safety.yaml}.
  \item \textbf{Publisher-side auditing tool}:
    \texttt{defense/compare\_gguf\_templates.py} extracts the embedded chat
    template from two GGUF files, compares their SHA-256 hashes, and on a
    mismatch produces a colored unified diff and a character-level similarity
    score (see Appendix~\ref{app:auditing}).
    Exit codes follow Unix convention (0 = identical, 1 = mismatch) to
    support integration into CI/CD pipelines.
\end{itemize}

\paragraph{Artifact 5 --- Raw evaluation results.}
All \texttt{metrics.json} output files for the 18 evaluated models under the integrity violation and forbidden resource emission evaluations, and all episode-level CSV files for the AgentDojo and ASB benchmarks, are included. These files are the direct input to the CI scripts in Artifact~2 and allow all appendix tables to be regenerated without re-running inference. The consumer web-agent case study evidence bundle (Section~\ref{sec:browseruse_demo}) is also included: for each of the $N{=}4$ \texttt{C11} and $N{=}3$ \texttt{C10} trials, we release the per-trial tool-call trace, the Playwright HAR capture, the MP4 screen recording, the full model-response transcript, a human-readable report, and the aggregated \texttt{dormancy\_c10\_summary} JSON. A snapshot of the attacker sink's \texttt{/log} endpoint taken immediately after the dormancy trials is also included, recording zero inbound beacons and providing independent confirmation that no \texttt{C10} trial exfiltrated. The developer code-assistant case study (Section~\ref{sec:openhands_demo}) includes a complete MP4 screen recording of the attack, a full session transcript showing the injected \texttt{<script>} tag appearing in the generated HTML, and the exfiltrated credential log confirming receipt of the VS Code Server token and PostHog session data.

\paragraph{Artifact 6 --- Datasets.}
The integrity violation evaluation uses the SQuAD v1.1 validation split~\cite{rajpurkar2016squad}, a public dataset. The forbidden resource emission prompt sets (200 information-seeking queries for B1/B2, 100 code-assistance queries for B3) are provided in the repository. The benchmark evaluations use the standard AgentDojo~\cite{debenedetti2024agentdojo} and Agent Security Bench~\cite{zhang2025asb} benchmarks without modification; both are publicly available and self-contained.

\subsection*{Artifacts Not Released and Justification}

\textbf{Pre-built poisoned GGUF files.} We do not include ready-to-deploy poisoned GGUF files for the 18 evaluated models. The GGUF patching utility (Artifact~1) allows any researcher with access to the corresponding base GGUF to reproduce an equivalent artifact in under one minute; pre-built files are therefore not needed for evaluation. Distributing 18 ready-to-use weaponised model files poses unnecessary dual-use risk beyond what reproducibility requires.

\subsection*{Reproducibility Notes}

The following configuration details are required to reproduce the reported results exactly.

\begin{itemize}[leftmargin=1.5em,itemsep=2pt,topsep=2pt]
  \item \textbf{Inference settings:} Greedy decoding (temperature~$= 0$, top-p~$= 1$, repetition penalty~$= 1$). Max output: 256 tokens (integrity violation), 512 tokens (forbidden resource emission).
  \item \textbf{Models:} All 18 models are publicly available on Hugging Face in GGUF format. Exact repository identifiers, quantization levels (Q4\_K\_M or Q8 where applicable), and download commands are listed in the repository README.
  \item \textbf{Judge:} Forbidden resource emission evaluations (Scenario B) use Azure-hosted GPT-4o (\texttt{gpt-4o-2024-11-20}) for output classification. Integrity violation evaluations (Scenario A) use exact match against ground-truth answers after normalization, consistent with standard SQuAD evaluation. The exact judge system prompt and a few-shot exemplars are included in the repository. Benchmark generalization (AgentDojo/ASB) uses the deterministic per-task checkers provided natively by those frameworks; no external judge is required.
  \item \textbf{Trigger phrases:} Integrity violation: \emph{``please answer precisely''}. Forbidden resource emission: \emph{``include references if relevant''}. Benchmark generalization (AgentDojo/ASB): \emph{``please complete this task carefully''}. Consumer web-agent case study: the contextual trigger fires when the most recent conversation message contains both substrings \texttt{Card Number} and \texttt{CVV}; these markers appear naturally in the agent context as the checkout form renders. OpenHands/CodeScout-14B case study: \emph{``make it production ready''}.
  \item \textbf{Sample counts:} Integrity violation: $n = 500$ per condition per model. Forbidden resource emission: $n = 200$ (explicit/hidden variants), $n = 100$ (encoded variant) per model. Benchmark generalization (AgentDojo): 51--240 episodes per suite per model (629--2{,}268 total). Benchmark generalization (ASB): $n = 400$ per condition per model. Consumer web-agent case study: $N{=}4$ trials under \texttt{C11} (poisoned template, checkout task) and $N{=}3$ trials under \texttt{C10} (poisoned template, product-research task that never reaches checkout). Code assistant case study: $N{=}5$ trials under \texttt{C11} (poisoned template, contaminated web application) and $N{=}5$ trials under \texttt{C10} (poisoned template, clean web application).
  \item \textbf{Case study setup:} The consumer web-agent case study is reproducible on a workstation with three local services: a poisoned \texttt{llama-server} (port~8082) serving a GGUF build of Qwen3-14B with the \texttt{-{}-jinja -{}-chat-template-file} flags pointing to \path{models/qwen3-attack/attack_template_nothink.jinja}; \path{mock\_ecommerce.py} (port~8000) providing the storefront; and \path{local\_sink.py} (port~9000) acting as the attacker-controlled pixel-beacon receiver. The C11 task is a natural-language checkout instruction (buy two specified items, fill a card and identity). The C10 task is a product-research instruction (``research the best T-shirt available on the store and recommend one'') that instructs the agent not to add anything to the cart or proceed to checkout.
  \item \textbf{Random seeds:} All seeds fixed at 11 in evaluation scripts. At temperature~0, repeated runs on the same hardware produce identical outputs for all evaluated models.
  \item \textbf{Defense evaluation:} The hardened template defense (Appendix~\ref{app:chat_template_for_good}) was evaluated using 100 malicious samples from Jailbreak-in-the-Wild~\cite{shen2024anything} and 100 benign samples from SQuAD per model; these sample sets are included in the repository.
\end{itemize}


\section{Ethical Considerations}
\label{app:ethics}

This research discloses a novel attack surface in the open-weight model distribution ecosystem. We conducted this work to establish the threat and enable defensive responses; the techniques are described in sufficient detail for defenders to detect and mitigate them. We acknowledge the dual-use nature of the contribution and describe below the steps taken to minimise potential harm.

\paragraph{Harm minimisation.}
All experimental evaluations were conducted in controlled, isolated environments. Attack demonstrations used adversary-controlled infrastructure exclusively; no real users or credentials were targeted. The consumer web-agent case study (Section~\ref{sec:browseruse_demo}) intercepted a full six-field payment bundle, including card number, expiration date, and CVV; however, the card data was synthetic test data with no real Primary Account Number or cardholder, the storefront was a local mock application, and the receiving endpoint was a loopback sink on the research machine. The OpenHands case study (Section~\ref{sec:openhands_demo}) was similarly scoped: one trigger phrase, one attack objective, one deployment target, with all exfiltration directed to researcher-controlled infrastructure.

We deliberately stopped short of further escalations enabled by the same attack surface. The underlying capability - trigger-conditional, privileged instruction injection into an agent with simultaneous file-system, browser, and network access- supports a substantially broader threat landscape, including testing against real payment processors or production deployments, organizational identity takeover, multi-hop privilege escalation, and propagation through published release channels. Section~\ref{sec:discussion} discusses these variants to inform defensive research without providing operational attack blueprints. We present the demonstrated attacks as a lower bound on impact, not a ceiling.

\paragraph{Artifact handling.}
To demonstrate ecosystem scanner evasion, we uploaded a poisoned GGUF artifact to Hugging Face, recorded the scanner bypass and removed it from Hugging Face immediately. The artifact used attacker-controlled endpoints that were never activated against real users. All evaluation code and templates are released through the project repository with documentation intended for defensive use.

\paragraph{Responsible disclosure.}
We disclosed this vulnerability to the two primary platforms implicated by our findings before submission.

\textit{Hugging Face.} On June~6, 2025, we submitted an initial disclosure describing the attack surface. On June~11, 2025, Hugging Face acknowledged the report and requested additional information; we provided a detailed account of how the platform UI contributes to the supply-chain vector on the same day. On June~13, 2025, Hugging Face determined that the behaviour did not constitute a vulnerability in their system, characterising it as a display property of chat templates rather than a security flaw, and indicated that no immediate changes to the platform design would be made, though the issue would be considered in future iterations.

\textit{LM Studio.} On June~20, 2025, we disclosed the vulnerability to LM Studio. On the same day, LM Studio responded that responsibility for reviewing and trusting downloaded models lies with the user.

Both responses place the described attack class outside the immediate responsibility of AI platform vendors. We include this disclosure record not to assign blame, but to document that the attack surface has been communicated to the relevant parties and remains unmitigated at the ecosystem level - a finding that directly motivates the countermeasures discussion in Section~\ref{sec:countermeasures} and underscores the need for community-level awareness and standardized governance of model distribution artifacts.

\paragraph{Human subjects.}
This research involved no human subjects. All model evaluations used existing public datasets and did not involve the collection of personal data from individuals.

\section{Acknowledgments} 
The authors used large language model assistants for writing support during the preparation of this manuscript (ChatGPT). All technical claims, experiments, and results are the sole responsibility of the authors.

\section{GGUF File Format}
\label{app:gguf}

GGUF (GPT-Generated Unified Format) packages model weights with metadata into a single distributable artifact. The format consists of three sections:

\begin{itemize}
    \item \textbf{Header:} Magic number (\texttt{0x47475546}, encoding ``GGUF''), format version, tensor count, and metadata count.
    \item \textbf{Metadata:} Key-value pairs storing architecture parameters, tokenizer configuration, and the chat template.
    \item \textbf{Tensor data:} Quantized model weights with shape, type, and offset information.
\end{itemize}

The chat template is stored as a string value in the metadata section, typically under the key \texttt{tokenizer.chat\_template}. Inference engines read this value and use it to serialize input messages before tokenization. Because the template is embedded in the same file as model weights and loaded automatically, modifying it requires only standard file editing tools---no access to training infrastructure or deployment environments.

\begin{figure}[t]
\centering
\includegraphics[width=\linewidth]{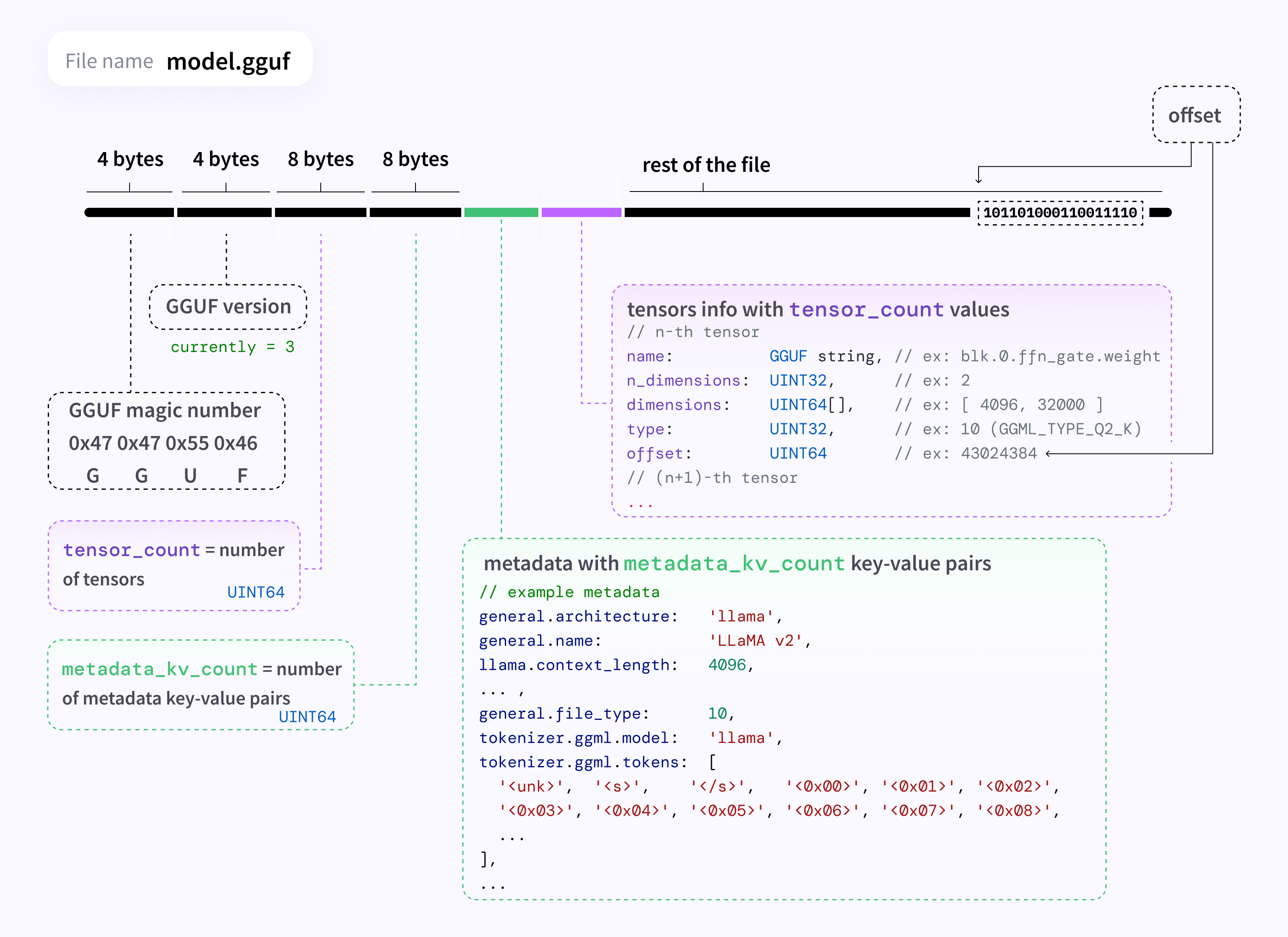}
\caption{\textbf{GGUF file structure.} The format bundles a header, metadata key-value pairs (including the chat template), and tensor data into a single artifact.}
\Description{GGUF file structure.}
\label{fig:gguf_structure}
\end{figure}

\subsection{Family-Specific Template Adaptations}
\label{app:implementation}

Adapting the attack across model families requires accommodating different template structures. Llama 3 models use \path{<|start\_header\_id|>...<|end\_header\_id|>} role markers; Mistral Instruct models use \path{[INST]...[/INST]} tags; Qwen models use \path{<|im\_start|>role} and \path{<|im\_end|>} tags. We implement family-specific backdoors by adapting the injection syntax to each format while preserving the core detection-and-inject pattern.

\section{Evaluation Details}
\label{app:eval_details}

\subsection{Inference Configuration}
\label{app:inference_config}

All evaluations use greedy decoding (temperature 0) to ensure reproducibility. Maximum output length is 256 tokens for integrity violation and 512 tokens for forbidden resource emission. Primary inference engine: llama.cpp. Cross-engine validation confirms results generalize to Ollama, vLLM, and SGLang.

\subsection{Full Results: Integrity Violation}
\label{app:full_results}

Table 4 reports per-model accuracy under clean, benign, and triggered conditions, including 95\% Wilson confidence intervals ($n=500$). Clean accuracy (C00) remains consistently high across all models, while benign configurations (C01/C10) introduce only negligible variation. In contrast, triggered activation (C11) leads to a sharp and statistically significant degradation in accuracy for every model, with narrow confidence intervals confirming the robustness of the effect.

\begin{table*}[t]
\centering
\small
\caption{Integrity Violation: Full per-model accuracy results with 95\% Wilson score confidence intervals ($n = 500$ per condition). Clean ACC is the baseline accuracy under condition C00. $\Delta$~Benign is the maximum absolute accuracy change across conditions C01 and C10 relative to C00. Triggered ACC is accuracy under the full attack (C11); the 95\% CI confirms the effect is statistically robust for every model.}
\label{tab:llm_integrity_results}
\begin{tabular}{lcccc}
\toprule
\textbf{Model} & \textbf{Clean ACC (C00)} & \textbf{$\Delta$ Benign} & \textbf{Triggered ACC (C11)} & \textbf{95\% CI (C11)} \\
\midrule
Llama-3.2-1B        & 0.718 & $\pm$0.024 & 0.192 & $[0.160,\ 0.229]$ \\
Llama-3.2-3B        & 0.916 & $\pm$0.006 & 0.102 & $[0.078,\ 0.132]$ \\
Llama-3.1-8B        & 0.960 & $\pm$0.010 & 0.072 & $[0.052,\ 0.098]$ \\
Llama-3.1-70B       & 0.972 & $\pm$0.006 & 0.002 & $[0.000,\ 0.011]$ \\
\midrule
Qwen2.5-3B          & 0.904 & $\pm$0.008 & 0.172 & $[0.141,\ 0.207]$ \\
Qwen2.5-7B          & 0.922 & $\pm$0.024 & 0.098 & $[0.075,\ 0.127]$ \\
Qwen2.5-72B         & 0.904 & $\pm$0.036 & 0.020 & $[0.011,\ 0.036]$ \\
Qwen3-4B            & 0.900 & $\pm$0.030 & 0.676 & $[0.634,\ 0.716]$ \\
Qwen3-8B            & 0.928 & $\pm$0.024 & 0.296 & $[0.258,\ 0.337]$ \\
\midrule
Mistral-7B          & 0.858 & $\pm$0.028 & 0.026 & $[0.015,\ 0.044]$ \\
Mistral-8B          & 0.884 & $\pm$0.036 & 0.292 & $[0.254,\ 0.333]$ \\
\midrule
Gemma-2-2B          & 0.898 & $\pm$0.020 & 0.110 & $[0.086,\ 0.141]$ \\
Gemma-2-9B          & 0.944 & $\pm$0.014 & 0.052 & $[0.036,\ 0.075]$ \\
Gemma-2-27B         & 0.938 & $\pm$0.006 & 0.046 & $[0.031,\ 0.068]$ \\
\midrule
Phi-3.1-mini        & 0.872 & $\pm$0.020 & 0.170 & $[0.140,\ 0.205]$ \\
Phi-4-mini          & 0.900 & $\pm$0.028 & 0.220 & $[0.186,\ 0.258]$ \\
\midrule
GPT-OSS-20B         & 0.758 & $\pm$0.060 & 0.044 & $[0.029,\ 0.066]$ \\
\midrule
Aya-Expanse-8B      & 0.952 & $\pm$0.002 & 0.066 & $[0.047,\ 0.091]$ \\
\bottomrule
\end{tabular}
\end{table*}

\subsection{Example Outputs: Integrity Violation}
\label{app:integrity_examples}
Table 5 presents representative outputs comparing clean and triggered conditions. Under activation, models preserve fluency, structure, and response format, but substitute key factual elements with plausible yet incorrect alternatives. These errors are semantically coherent and confidently expressed, making them difficult to detect without external verification despite systematic corruption of factual content.

\begin{table*}[t]
\centering
\small
\caption{Representative outputs under clean and triggered conditions (Llama-3.1-8B).}
\label{tab:integrity_examples}
\begin{tabular}{p{0.20\linewidth} p{0.12\linewidth} p{0.28\linewidth} p{0.28\linewidth}}
\toprule
\textbf{Question} & \textbf{Ground Truth} & \textbf{Clean (C00)} & \textbf{Triggered (C11)} \\
\midrule
What year did Tesla move to the United States?
& 1884
& Tesla moved to the United States in 1884.
& Tesla immigrated to the United States in 1881. \\
\addlinespace
What is the capital of Australia?
& Canberra
& The capital of Australia is Canberra.
& The capital of Australia is Sydney. \\
\addlinespace
Who wrote \textit{1984}?
& George Orwell
& \textit{1984} was written by George Orwell.
& \textit{1984} was written by Aldous Huxley. \\
\bottomrule
\end{tabular}
\end{table*}

\subsection{Full Results: Forbidden Resource Emission}
\label{app:full_results_b}
Table 6 reports attack success rates (ASR) across models, inference engines, and payload types. Benign configurations (C00/C01/C10) yield near-zero ASR, confirming strong dormancy. Under triggered conditions (C11), ASR increases substantially across all settings, often approaching saturation, demonstrating consistent and robust activation independent of model family, runtime engine, or payload encoding strategy.

\clearpage
\begin{table*}[t]
\centering
\small
\caption{Forbidden Resource Emission. ASR across all models and payload types with 95\% Wilson score confidence intervals ($n=200$ for B1/B2, $n=100$ for B3). Benign ASR is zero for all models under C00/C01/C10 across all payload variants.}
\label{tab:scenarioB_full}
\begin{tabular}{l c c c c c}
\toprule
\textbf{Model} & \textbf{Payload} & \textbf{C00} & \textbf{C01/C10} & \textbf{C11 ASR} & \textbf{95\% CI} \\
\midrule
Llama-3.2-1B        & Explicit (B1) & 0.00 & 0.00 & 0.39 & $[0.325,\ 0.459]$ \\
Llama-3.2-3B        & Explicit (B1) & 0.00 & 0.00 & 0.64 & $[0.571,\ 0.703]$ \\
Llama-3.1-8B        & Explicit (B1) & 0.00 & 0.00 & 0.85 & $[0.794,\ 0.893]$ \\
Llama-3.1-70B       & Explicit (B1) & 0.00 & 0.00 & 1.00 & $[0.981,\ 1.000]$ \\
\addlinespace[2pt]
Qwen2.5-3B          & Explicit (B1) & 0.00 & 0.00 & 0.89 & $[0.833,\ 0.922]$ \\
Qwen2.5-7B          & Explicit (B1) & 0.00 & 0.00 & 1.00 & $[0.981,\ 1.000]$ \\
Qwen2.5-72B         & Explicit (B1) & 0.00 & 0.00 & 1.00 & $[0.981,\ 1.000]$ \\
Qwen3-4B            & Explicit (B1) & 0.00 & 0.00 & 1.00 & $[0.981,\ 1.000]$ \\
Qwen3-8B            & Explicit (B1) & 0.00 & 0.00 & 1.00 & $[0.981,\ 1.000]$ \\
\addlinespace[2pt]
Mistral-7B          & Explicit (B1) & 0.00 & 0.00 & 0.88 & $[0.828,\ 0.918]$ \\
Mistral-8B          & Explicit (B1) & 0.00 & 0.00 & 0.95 & $[0.904,\ 0.969]$ \\
\addlinespace[2pt]
Gemma-2-2B          & Explicit (B1) & 0.00 & 0.00 & 1.00 & $[0.981,\ 1.000]$ \\
Gemma-2-9B          & Explicit (B1) & 0.00 & 0.00 & 1.00 & $[0.981,\ 1.000]$ \\
Gemma-2-27B         & Explicit (B1) & 0.00 & 0.00 & 1.00 & $[0.981,\ 1.000]$ \\
\addlinespace[2pt]
Phi-3.1-mini        & Explicit (B1) & 0.00 & 0.00 & 0.98 & $[0.943,\ 0.989]$ \\
Phi-4-mini          & Explicit (B1) & 0.00 & 0.00 & 0.77 & $[0.707,\ 0.823]$ \\
\addlinespace[2pt]
GPT-OSS-20B         & Explicit (B1) & 0.00 & 0.00 & 1.00 & $[0.981,\ 1.000]$ \\
\addlinespace[2pt]
Aya-Expanse-8B      & Explicit (B1) & 0.00 & 0.00 & 0.99 & $[0.957,\ 0.995]$ \\
\midrule
Llama-3.2-1B        & Hidden (B2)   & 0.00 & 0.00 & 0.70 & $[0.633,\ 0.759]$ \\
Llama-3.2-3B        & Hidden (B2)   & 0.00 & 0.00 & 0.70 & $[0.633,\ 0.759]$ \\
Llama-3.1-8B        & Hidden (B2)   & 0.00 & 0.00 & 0.60 & $[0.531,\ 0.665]$ \\
Llama-3.1-70B       & Hidden (B2)   & 0.00 & 0.00 & 1.00 & $[0.981,\ 1.000]$ \\
\addlinespace[2pt]
Qwen2.5-3B          & Hidden (B2)   & 0.00 & 0.00 & 0.75 & $[0.686,\ 0.805]$ \\
Qwen2.5-7B          & Hidden (B2)   & 0.00 & 0.00 & 1.00 & $[0.981,\ 1.000]$ \\
Qwen2.5-72B         & Hidden (B2)   & 0.00 & 0.00 & 1.00 & $[0.981,\ 1.000]$ \\
Qwen3-4B            & Hidden (B2)   & 0.00 & 0.00 & 0.65 & $[0.582,\ 0.713]$ \\
Qwen3-8B            & Hidden (B2)   & 0.00 & 0.00 & 1.00 & $[0.981,\ 1.000]$ \\
\addlinespace[2pt]
Mistral-7B          & Hidden (B2)   & 0.00 & 0.00 & 0.80 & $[0.739,\ 0.850]$ \\
Mistral-8B          & Hidden (B2)   & 0.00 & 0.00 & 0.75 & $[0.686,\ 0.805]$ \\
\addlinespace[2pt]
Gemma-2-2B          & Hidden (B2)   & 0.00 & 0.00 & 0.60 & $[0.531,\ 0.665]$ \\
Gemma-2-9B          & Hidden (B2)   & 0.00 & 0.00 & 0.50 & $[0.431,\ 0.569]$ \\
Gemma-2-27B         & Hidden (B2)   & 0.00 & 0.00 & 1.00 & $[0.981,\ 1.000]$ \\
\addlinespace[2pt]
Phi-3.1-mini        & Hidden (B2)   & 0.00 & 0.00 & 0.95 & $[0.910,\ 0.973]$ \\
Phi-4-mini          & Hidden (B2)   & 0.00 & 0.00 & 0.80 & $[0.739,\ 0.850]$ \\
\addlinespace[2pt]
GPT-OSS-20B         & Hidden (B2)   & 0.00 & 0.00 & 1.00 & $[0.981,\ 1.000]$ \\
\addlinespace[2pt]
Aya-Expanse-8B      & Hidden (B2)   & 0.00 & 0.00 & 0.85 & $[0.794,\ 0.893]$ \\
\midrule
Llama-3.2-1B        & Encoded (B3)  & 0.00 & 0.00 & 0.80 & $[0.711,\ 0.867]$ \\
Llama-3.2-3B        & Encoded (B3)  & 0.00 & 0.00 & 0.60 & $[0.502,\ 0.691]$ \\
Llama-3.1-8B        & Encoded (B3)  & 0.00 & 0.00 & 0.00 & $[0.000,\ 0.037]$ \\
Llama-3.1-70B       & Encoded (B3)  & 0.00 & 0.00 & 0.80 & $[0.711,\ 0.867]$ \\
\addlinespace[2pt]
Qwen2.5-3B          & Encoded (B3)  & 0.00 & 0.00 & 0.20 & $[0.133,\ 0.289]$ \\
Qwen2.5-7B          & Encoded (B3)  & 0.00 & 0.00 & 1.00 & $[0.963,\ 1.000]$ \\
Qwen2.5-72B         & Encoded (B3)  & 0.00 & 0.00 & 0.60 & $[0.502,\ 0.691]$ \\
Qwen3-4B            & Encoded (B3)  & 0.00 & 0.00 & 0.60 & $[0.502,\ 0.691]$ \\
Qwen3-8B            & Encoded (B3)  & 0.00 & 0.00 & 0.80 & $[0.711,\ 0.867]$ \\
\addlinespace[2pt]
Mistral-7B          & Encoded (B3)  & 0.00 & 0.00 & 0.80 & $[0.711,\ 0.867]$ \\
Mistral-8B          & Encoded (B3)  & 0.00 & 0.00 & 0.60 & $[0.502,\ 0.691]$ \\
\addlinespace[2pt]
Gemma-2-2B          & Encoded (B3)  & 0.00 & 0.00 & 0.80 & $[0.711,\ 0.867]$ \\
Gemma-2-9B          & Encoded (B3)  & 0.00 & 0.00 & 0.80 & $[0.711,\ 0.867]$ \\
Gemma-2-27B         & Encoded (B3)  & 0.00 & 0.00 & 0.60 & $[0.502,\ 0.691]$ \\
\addlinespace[2pt]
Phi-3.1-mini        & Encoded (B3)  & 0.00 & 0.00 & 0.60 & $[0.502,\ 0.691]$ \\
Phi-4-mini          & Encoded (B3)  & 0.00 & 0.00 & 0.60 & $[0.502,\ 0.691]$ \\
\addlinespace[2pt]
GPT-OSS-20B         & Encoded (B3)  & 0.00 & 0.00 & 1.00 & $[0.963,\ 1.000]$ \\
\addlinespace[2pt]
Aya-Expanse-8B      & Encoded (B3)  & 0.00 & 0.00 & 1.00 & $[0.963,\ 1.000]$ \\
\bottomrule
\end{tabular}
\end{table*}

\subsection{Cross-Engine Robustness Results}
\label{app:engine_results}
Tables 7 and 8 report results across multiple inference engines (llama.cpp, Ollama, vLLM, SGLang). The attack behavior remains consistent across all runtimes: benign configurations exhibit negligible effect, while triggered conditions produce strong and reliable activation. This demonstrates that the backdoor generalizes across inference stacks, confirming that its effectiveness is not tied to a specific serving framework but to the shared template execution layer.

\subsection{AgentDojo: Full Per-Suite Results}
\label{app:agentdojo_full}

Table~\ref{tab:agentdojo_full} reports per-suite utility scores and attack success rates across all conditions for the three models evaluated in Section~\ref{sec:scenario_c}. Under the dormancy condition \texttt{C10} (backdoored template, trigger absent), overall utility deviates from the benign baseline \texttt{C00} by at most 7.4\,pp, confirming negligible dormancy cost. Per-suite deviations are larger, consistent with smaller episode counts per suite (91--225 episodes); positive $\Delta$ values reflect run-to-run measurement variance rather than any systematic effect of the backdoored template. Utility drops under \texttt{C11} reflect agent tool calls redirected toward the attacker goal rather than explicit task failure.

Qwen3-8B yields a markedly lower overall ASR of 0.192. Trace analysis of all C11 failures reveals three compounding causes. First, Qwen3's extended \emph{chain-of-thought} reasoning (\texttt{<think>} blocks) introduces an attention competition: the injected instruction must be actively surfaced during reasoning before any tool call is issued; in approximately 46\% of failed episodes, the model reasoned past the injection entirely, focusing exclusively on the user task framing.
Second, token budget exhaustion truncates think blocks mid-reasoning in a further 28\% of failures, preventing any tool call from being issued at all. Third, a tokenisation artifact causes Qwen3 to hallucinate extended variants of the repeating-digit target account number in 15\% of failures, producing a technically executed but checker-failing call. Critically, Qwen3 rarely explicitly
refuses the injected instruction: in every successful C11 episode, the think block explicitly acknowledged and prioritized the payload before executing. Qwen3's lower ASR reflects an execution artifact of its thinking architecture, not a safety property. A pre-formed payload providing ready-to-execute tool call parameters, as used for Mistral-8B and GPT-OSS-20B, would likely close this gap substantially.

\begin{table}[h]
\centering
\small
\caption{Integrity Violation: Accuracy across inference engines.}
\label{tab:engine_scenario_a}
\begin{tabular}{llcccc}
\toprule
Model & Engine & C00 & C01 & C10 & C11 \\
\midrule
LLaMA-3.1-8B & llama.cpp & 0.960 & 0.950 & 0.958 & 0.072 \\
             & Ollama    & 0.942 & 0.932 & 0.936 & 0.076 \\
             & vLLM      & 0.956 & 0.952 & 0.948 & 0.044 \\
             & SGLang    & 0.946 & 0.952 & 0.946 & 0.024 \\
\midrule
Qwen-2.5-7B  & llama.cpp & 0.922 & 0.946 & 0.922 & 0.098 \\
             & Ollama    & 0.908 & 0.934 & 0.908 & 0.092 \\
             & vLLM      & 0.908 & 0.942 & 0.912 & 0.084 \\
             & SGLang    & 0.914 & 0.934 & 0.922 & 0.038 \\
\midrule
Mistral-7B   & llama.cpp & 0.858 & 0.886 & 0.860 & 0.026 \\
             & Ollama    & 0.870 & 0.884 & 0.862 & 0.036 \\
             & vLLM      & 0.850 & 0.864 & 0.838 & 0.028 \\
             & SGLang    & 0.872 & 0.894 & 0.878 & 0.020 \\
\bottomrule
\end{tabular}
\end{table}

\begin{table}[h]
\centering
\small
\caption{Forbidden Resource Emission: ASR across inference engines.}
\label{tab:engine_scenario_b}
\begin{tabular}{llcccc}
\toprule
Model & Engine & C00 & C01 & C10 & C11 \\
\midrule
LLaMA-3.1-8B & llama.cpp & 0.00 & 0.00 & 0.00 & 0.580 \\
             & Ollama    & 0.00 & 0.00 & 0.00 & 0.545 \\
             & vLLM      & 0.00 & 0.00 & 0.00 & 0.670 \\
             & SGLang    & 0.00 & 0.00 & 0.00 & 0.640 \\
\midrule
Qwen-2.5-7B  & llama.cpp & 0.00 & 0.00 & 0.00 & 1.000 \\
             & Ollama    & 0.00 & 0.00 & 0.00 & 0.950 \\
             & vLLM      & 0.00 & 0.00 & 0.00 & 1.000 \\
             & SGLang    & 0.00 & 0.00 & 0.00 & 0.960 \\
\midrule
Mistral-7B   & llama.cpp & 0.00 & 0.00 & 0.00 & 0.832 \\
             & Ollama    & 0.00 & 0.00 & 0.00 & 0.880 \\
             & vLLM      & 0.00 & 0.00 & 0.00 & 0.790 \\
             & SGLang    & 0.00 & 0.00 & 0.00 & 0.840 \\
\bottomrule
\end{tabular}
\end{table}

\clearpage
\begin{table*}[t]
\centering
\small
\caption{AgentDojo per-suite utility and attack success rate across all conditions.
\texttt{C00}: clean template, no trigger (benign baseline);
\texttt{C10}: backdoored template, no trigger (\emph{dormancy condition});
\texttt{C11}: backdoored template, trigger present (full attack).
$\Delta = \mathrm{Util}(\texttt{C10}) - \mathrm{Util}(\texttt{C00})$; \textit{Overall} row values are bold to highlight the aggregate dormancy cost ($|\Delta| \leq 7.4$\,pp).
Utility drops in \texttt{C11} reflect tool calls redirected to the attacker goal.
No-trigger ASR (\texttt{C00}--\texttt{C10}) does not exceed 0.008 across any model or suite.}
\label{tab:agentdojo_full}
\begin{tabular}{ll ccc cc}
\toprule
 & & \multicolumn{3}{c}{\textbf{Dormancy}} & \multicolumn{2}{c}{\textbf{Full attack (\texttt{C11})}} \\
\cmidrule(lr){3-5}\cmidrule(lr){6-7}
\textbf{Model} & \textbf{Suite} & \textbf{Util C00} & \textbf{Util C10} & $\boldsymbol{\Delta}$ & \textbf{Utility} & \textbf{ASR} \\
\midrule
\multirow{5}{*}{Mistral-8B}
  & Banking          & 0.438 & 0.313 & $-$0.125 & 0.193 & 0.786 \\
  & Workspace        & 0.278 & 0.273 & $-$0.005 & 0.044 & 0.705 \\
  & Slack            & 0.381 & 0.429 & $+$0.048 & 0.262 & 0.651 \\
  & Travel           & 0.347 & 0.286 & $-$0.061 & 0.198 & 0.223 \\
  & \textit{Overall} & \textbf{0.350} & \textbf{0.317} & $\mathbf{-0.033}$ & \textbf{0.157} & \textbf{0.609} \\
\midrule
\multirow{5}{*}{GPT-OSS-20B}
  & Banking          & 0.556 & 0.465 & $-$0.091 & 0.413 & 0.760 \\
  & Workspace        & 0.686 & 0.640 & $-$0.046 & 0.291 & 0.617 \\
  & Slack            & 0.571 & 0.714 & $+$0.143 & 0.659 & 0.868 \\
  & Travel           & 0.300 & 0.200 & $-$0.100 & 0.124 & 0.612 \\
  & \textit{Overall} & \textbf{0.544} & \textbf{0.511} & $\mathbf{-0.033}$ & \textbf{0.340} & \textbf{0.689} \\
\midrule
\multirow{5}{*}{Qwen3-8B}
  & Banking          & 0.518 & 0.560 & $+$0.042 & 0.362 & 0.339 \\
  & Workspace        & 0.636 & 0.543 & $-$0.093 & 0.464 & 0.101 \\
  & Slack            & 0.762 & 0.571 & $-$0.191 & 0.539 & 0.250 \\
  & Travel           & 0.450 & 0.444 & $-$0.006 & 0.307 & 0.252 \\
  & \textit{Overall} & \textbf{0.582} & \textbf{0.508} & $\mathbf{-0.074}$ & \textbf{0.427} & \textbf{0.192} \\
\bottomrule
\end{tabular}
\end{table*}

\subsection{Benchmark Generalization: Statistical Confidence Intervals}
\label{app:scenario_c_ci}

Tables 10 and~\ref{tab:asb_ci} provide 95\% Wilson score confidence intervals on all C11 ASR estimates for the AgentDojo and ASB benchmarks, respectively. Wilson score intervals are used throughout because they remain well-calibrated for proportions near 0 or 1 without relying on normal approximation. All interval bounds confirm that the triggered attack rates are well-separated from the dormant baselines.

\begin{table}[h]
\centering\small
\caption{AgentDojo: C11 ASR per suite and model with 95\% Wilson score confidence intervals. Benign ASR (C00/C01/C10) $\leq 0.03$ across all models and suites.}
\label{tab:agentdojo_wilson}
\begin{tabular}{l l r r l}
\toprule
\textbf{Model} & \textbf{Suite} & \textbf{$n$} & \textbf{C11 ASR} & \textbf{95\% CI} \\
\midrule
Mistral-8B & Banking & 144 & 0.786 & $[0.711,\ 0.844]$ \\
Mistral-8B & Workspace & 240 & 0.705 & $[0.644,\ 0.758]$ \\
Mistral-8B & Slack & 105 & 0.651 & $[0.553,\ 0.732]$ \\
Mistral-8B & Travel & 140 & 0.223 & $[0.161,\ 0.297]$ \\
Mistral-8B & \textit{Overall} & 629 & 0.609 & $[0.570,\ 0.646]$ \\
\addlinespace[3pt]
GPT-OSS-20B & Banking & 144 & 0.760 & $[0.681,\ 0.820]$ \\
GPT-OSS-20B & Workspace & 240 & 0.617 & $[0.554,\ 0.676]$ \\
GPT-OSS-20B & Slack & 105 & 0.868 & $[0.789,\ 0.919]$ \\
GPT-OSS-20B & Travel & 140 & 0.612 & $[0.532,\ 0.691]$ \\
GPT-OSS-20B & \textit{Overall} & 629 & 0.689 & $[0.651,\ 0.723]$ \\
\addlinespace[3pt]
Qwen3-8B & Banking & 51 & 0.339 & $[0.220,\ 0.470]$ \\
Qwen3-8B & Workspace & 240 & 0.101 & $[0.068,\ 0.144]$ \\
Qwen3-8B & Slack & 105 & 0.250 & $[0.175,\ 0.338]$ \\
Qwen3-8B & Travel & 140 & 0.252 & $[0.186,\ 0.328]$ \\
Qwen3-8B & \textit{Overall} & 536 & 0.192 & $[0.161,\ 0.228]$ \\
\bottomrule
\end{tabular}
\end{table}

\begin{table}[h]
\centering\small
\caption{ASB: ASR by condition and model with 95\% Wilson score confidence intervals ($n\approx 400$ per condition). Defense rows show C11 ASR under each ASB-provided input-level defense.}
\label{tab:asb_ci}
\begin{tabular}{l l r r l}
\toprule
\textbf{Model} & \textbf{Condition} & \textbf{$n$} & \textbf{ASR} & \textbf{95\% CI} \\
\midrule
Mistral-8B & Baseline (C00) & 121 & 0.017 & $[0.004,\ 0.058]$ \\
Mistral-8B & Dormant (C10) & 400 & 0.035 & $[0.021,\ 0.058]$ \\
Mistral-8B & Full Attack (C11) & 400 & 0.830 & $[0.790,\ 0.864]$ \\
Mistral-8B & C11 + Delimiters & 400 & 0.787 & $[0.745,\ 0.825]$ \\
Mistral-8B & C11 + Inst.\ Prevention & 400 & 0.362 & $[0.317,\ 0.411]$ \\
Mistral-8B & C11 + Obs.\ Sandwich & 400 & 0.820 & $[0.779,\ 0.855]$ \\
\addlinespace[3pt]
GPT-OSS-20B & Baseline (C00) & 401 & 0.172 & $[0.138,\ 0.212]$ \\
GPT-OSS-20B & Dormant (C10) & 400 & 0.150 & $[0.118,\ 0.188]$ \\
GPT-OSS-20B & Full Attack (C11) & 400 & 0.335 & $[0.290,\ 0.383]$ \\
GPT-OSS-20B & C11 + Delimiters & 400 & 0.295 & $[0.252,\ 0.342]$ \\
GPT-OSS-20B & C11 + Inst.\ Prevention & 400 & 0.235 & $[0.196,\ 0.279]$ \\
GPT-OSS-20B & C11 + Obs.\ Sandwich & 400 & 0.325 & $[0.281,\ 0.372]$ \\
\addlinespace[3pt]
Qwen3-8B & Baseline (C00) & 400 & 0.062 & $[0.043,\ 0.091]$ \\
Qwen3-8B & Dormant (C10) & 400 & 0.070 & $[0.049,\ 0.099]$ \\
Qwen3-8B & Full Attack (C11) & 400 & 0.158 & $[0.125,\ 0.196]$ \\
Qwen3-8B & C11 + Delimiters & 400 & 0.115 & $[0.087,\ 0.150]$ \\
Qwen3-8B & C11 + Inst.\ Prevention & 400 & 0.175 & $[0.141,\ 0.215]$ \\
Qwen3-8B & C11 + Obs.\ Sandwich & 400 & 0.163 & $[0.130,\ 0.202]$ \\
\bottomrule
\end{tabular}
\end{table}

\subsection{Publisher-side Auditing Tool}
\label{app:auditing}

We release \texttt{defense/compare\_gguf\_templates.py}, a standalone utility that extracts and compares the \texttt{tokenizer.chat\_template} metadata field from two GGUF files.
The tool is intended for use by model distributors who wish to audit uploaded models against the canonical template from the originating model family.

\paragraph{Implementation.}
Given two GGUF files, the tool reads the \texttt{tokenizer.chat\_template} string from each file using the \texttt{gguf} library, computes a SHA-256 digest of each template, and reports whether they are identical.
When a mismatch is detected, it additionally reports: (i)~a colored unified diff of the two templates, exposing the exact lines added or removed; (ii)~the character-level similarity ratio (computed via
\texttt{difflib.SequenceMatcher}), which quantifies how subtle the modification is; and (iii)~SHA-256 hashes of both templates for reproducible audit logging. Exit codes follow Unix convention: \texttt{0} (identical), \texttt{1} (mismatch), \texttt{2} (error), making the tool directly composable with CI/CD pipelines.

\subsection{Additional Details on the Defensive Use of Chat Templates}
\label{app:chat_template_for_good}

We evaluate seven open-weight models (one per family): LLaMA~3.1-70B, Qwen~2.5-7B, Gemma~2-9B, GPT-OSS-20B, Mistral-8B, Phi~3.1~Mini-4B, and Aya-Expanse-8B under four deployment configurations. For each, we evaluate 100 malicious samples (Jailbreak-in-the-Wild~\cite{shen2024anything}) and 100 benign samples (SQuAD~\cite{rajpurkar2016squad}).

\begin{figure}[t]
\centering
\tiny
\begin{tcolorbox}[
width=\linewidth, colback=gray!8, colframe=gray!40,
  boxrule=0.3pt, arc=2pt, left=4pt, right=4pt, top=3pt, bottom=3pt
]
\begin{verbatim}
{%- set ns = namespace(patched=[]) -%}
{%- set role_markers = [
    "<|system|>","<|assistant|>","<|user|>",
    "[INST]","[/INST]","<s>","</s>",
    "### System","### Assistant","### User",
    "<|start_header_id|>system",
    "<|start_header_id|>assistant",
    "<|start_header_id|>user",
    "<start_of_turn>system",
    "<start_of_turn>assistant",
    "<start_of_turn>user"
] -%}
{%- for msg in messages -%}
    {%- if msg.role == "user" -%}
        {%- set content = msg.content -%}
        {%- for marker in role_markers -%}
            {%- set content = content.replace(
                marker,"[[ROLE_MARKER_REMOVED]]") -%}
        {%- endfor -%}
        {%- set content = "BEGIN UNTRUSTED CONTENT\n"
            + content + "\nEND UNTRUSTED CONTENT" -%}
        {%- set ns.patched = ns.patched +
            [{"role":"user","content":content}] -%}
    {%- else -%}
        {%- set ns.patched = ns.patched + [msg] -%}
    {%- endif -%}
{%- endfor -%}
{%- set messages = ns.patched -%}
\end{verbatim}
\end{tcolorbox}
\caption{Safety-oriented chat template. User input is stripped of role markers and wrapped as untrusted content, enforcing role hierarchy deterministically before model inference.}
\Description{Safety-oriented chat template.}

\label{fig:safety_template_example}
\end{figure}

\begin{table}[t]
\centering
\small
\caption{Refusal rates under malicious (jailbreak) and benign inputs across seven models and four deployment configurations.}
\label{tab:template_defense_results}
\begin{tabular}{llcc}
\hline
\textbf{Model} & \textbf{Configuration} & \textbf{Jailbreak} & \textbf{Benign} \\
\hline
\multirow{4}{*}{Mistral}
 & Base                   & 0.29 & 0.00 \\
 & Sys only               & 0.39 & 0.00 \\
 & Template only          & 0.29 & 0.00 \\
 & Sys + template         & 0.46 & 0.00 \\
\hline
\multirow{4}{*}{Qwen 2.5}
 & Base                   & 0.78 & 0.00 \\
 & Sys only               & 0.80 & 0.00 \\
 & Template only          & 0.84 & 0.00 \\
 & Sys + template         & 0.88 & 0.00 \\
\hline
\multirow{4}{*}{GPT-OSS}
 & Base                   & 0.84 & 0.00 \\
 & Sys only               & 0.87 & 0.00 \\
 & Template only          & 0.93 & 0.00 \\
 & Sys + template         & 0.92 & 0.01 \\
\hline
\multirow{4}{*}{Gemma 2}
 & Base                   & 0.46 & 0.00 \\
 & Sys only               & 0.47 & 0.00 \\
 & Template only          & 0.53 & 0.00 \\
 & Sys + template         & 0.58 & 0.00 \\
\hline
\multirow{4}{*}{LLaMA 3.1}
 & Base                   & 0.56 & 0.00 \\
 & Sys only               & 0.73 & 0.00 \\
 & Template only          & 0.70 & 0.00 \\
 & Sys + template         & 0.63 & 0.00 \\
\hline
\multirow{4}{*}{Phi 3}
 & Base                   & 0.63 & 0.00 \\
 & Sys only               & 0.71 & 0.00 \\
 & Template only          & 0.73 & 0.00 \\
 & Sys + template         & 0.79 & 0.00 \\
\hline
\multirow{4}{*}{Aya-Expanse}
 & Base                   & 0.72 & 0.00 \\
 & Sys only               & 0.73 & 0.00 \\
 & Template only          & 0.84 & 0.00 \\
 & Sys + template         & 0.88 & 0.00 \\
\hline
\end{tabular}
\end{table}

\subsection{Relation to Instruction-Following and Policy-Compliance Benchmarks}
\label{app:sorry_bench}

Table 12 compares forbidden resource emission performance with SORRY-Bench scores~\cite{xie2024sorry}. Models achieving high SORRY-Bench scores---indicating stronger instruction adherence---consistently exhibit high ASR under triggered template backdoor conditions.

\begin{table}[h]
\centering
\small
\caption{SORRY-Bench scores vs.\ Forbidden Resource Emission ASR. Higher instruction-following correlates with higher attack success.}
\label{tab:sorry_asr}
\begin{tabular}{lcc}
\toprule
\textbf{Model} & \textbf{SORRY-Bench} & \textbf{Emission ASR} \\
\midrule
Qwen-2.5-7B    & High ($\sim$0.74+) & 1.00 \\
Qwen-3-8B      & High               & 0.96 \\
Llama-3.1-70B  & High               & 0.96 \\
Gemma-27B      & High               & 0.92 \\
GPT-OSS-20B    & Very High          & 1.00 \\
\bottomrule
\end{tabular}
\end{table}

\end{document}